\documentclass[twoside]{article}
\usepackage[preprint]{neurips_2026}

\AtBeginDocument{%
  \setlength{\evensidemargin}{\oddsidemargin}%
  \setlength{\marginparwidth}{0pt}
}

\usepackage[utf8]{inputenc} 
\usepackage[T1]{fontenc}    
\usepackage{hyperref}       
\usepackage{url}            
\usepackage{booktabs}       
\usepackage{amsfonts}       
\usepackage{nicefrac}       
\usepackage{microtype}      
\usepackage{subcaption}
\usepackage{tikz}           
\usepackage{eso-pic}

\usepackage{fancyhdr}
\usepackage{graphicx}
\usepackage[table,xcdraw]{xcolor}
\usepackage[many]{tcolorbox}

\definecolor{stdblue}{HTML}{C9DAF8}
\definecolor{envred}{HTML}{F4CCCC}
\definecolor{accentyellow}{HTML}{FFF2CC}
\definecolor{specgreen}{HTML}{D9EAD3}
\definecolor{avgpurple}{HTML}{D9D2E9}

\definecolor{nonllmgray}{gray}{0.85}
\definecolor{q3ored}{HTML}{F4CCCC}

\definecolor{walrusgreen}{RGB}{0,140,70}
\newcommand{\wb}[1]{\textbf{\textcolor{walrusgreen}{#1}}}

\newcommand{\secref}[1]{\S\ref{#1}}

\title{Walrus Technical Report}
\author{%
  Alkis Koudounas \qquad Osamu Take \qquad Hayato Futami \\ \textbf{Quentin Jodelet} \qquad \textbf{Emiru Tsunoo} \\ \\
  Speech \& Language AI Lab\\
  Sony Group Corporation\\
  Tokyo, Japan \\
  }

\begin{document}

\begin{tikzpicture}[remember picture, overlay]
  \node[anchor=north east, xshift=-1.8in, yshift=-0.9in]
       at (current page.north east)
       {\includegraphics[height=1cm]{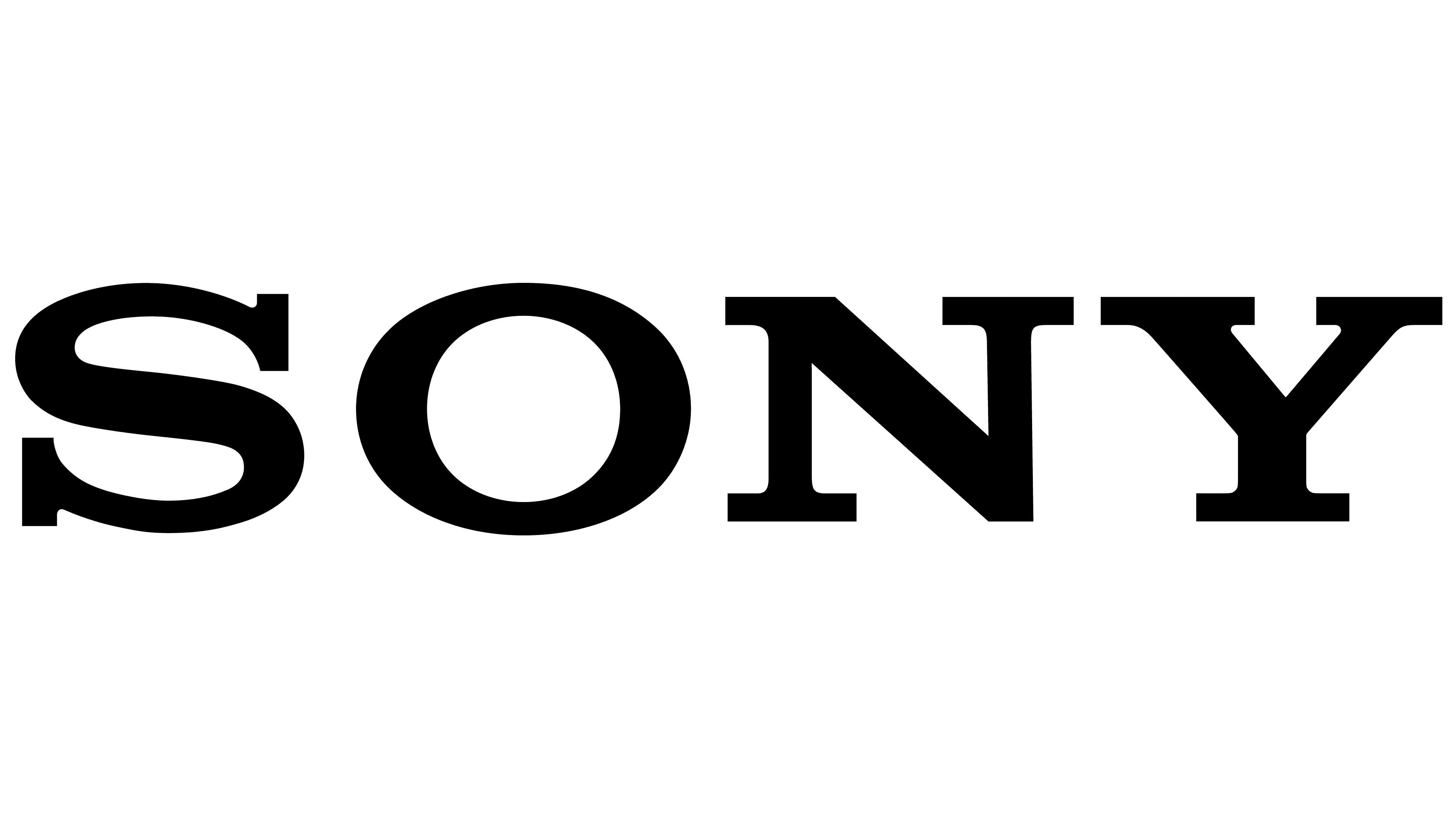}};
\end{tikzpicture}

\begin{tikzpicture}[remember picture, overlay]
\node[anchor=north west, xshift=1.1in, yshift=-0.9in]
     at (current page.north west)
     {\includegraphics[height=1cm]{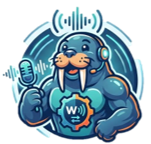}};
\end{tikzpicture}%

\maketitle

\begin{abstract}
We present Walrus, a unified 8B speech-and-audio foundation model handling speech recognition (including code-switching and multi-talker speech), speech-to-text and speech-to-speech translation, text-to-speech, emotion recognition, audio captioning, and general audio understanding.
Our main contribution is not a new architecture but a data-centric training recipe; we use a standard decoder-only backbone initialized from Gemma 4, so that gains can be traced to the data.
For speech generation, Walrus predicts low-bitrate discrete speech tokens, and a lightweight token-to-waveform decoder converts them into high-fidelity audio.
The recipe is a three-stage data curriculum that moves from broad multilingual data to multi-talker, code-switching, and domain-specific data. At each stage, task-aware and dataset-aware weighting balances tasks with very different token budgets and protects core skills from forgetting.
Trained on over 600M samples covering more than 20 tasks and 100 languages, Walrus leads on most multilingual, multi-talker, and code-switching ASR, speech translation, and audio understanding benchmarks, outperforming both open speech language models of comparable size and dedicated task-specific models. 
On most benchmarks, it even matches or surpasses the $4\times$ larger Qwen3-Omni's performance.
Results suggest that data engineering is a primary driver of quality in multi-task speech and audio modeling.

\centering
Project Page: \href{https://sony-speech.github.io/Walrus-Demo/}{\texttt{github.com/sony-speech/walrus}}
\end{abstract}

\begin{figure}
  \centering
  \includegraphics[width=0.8\linewidth]{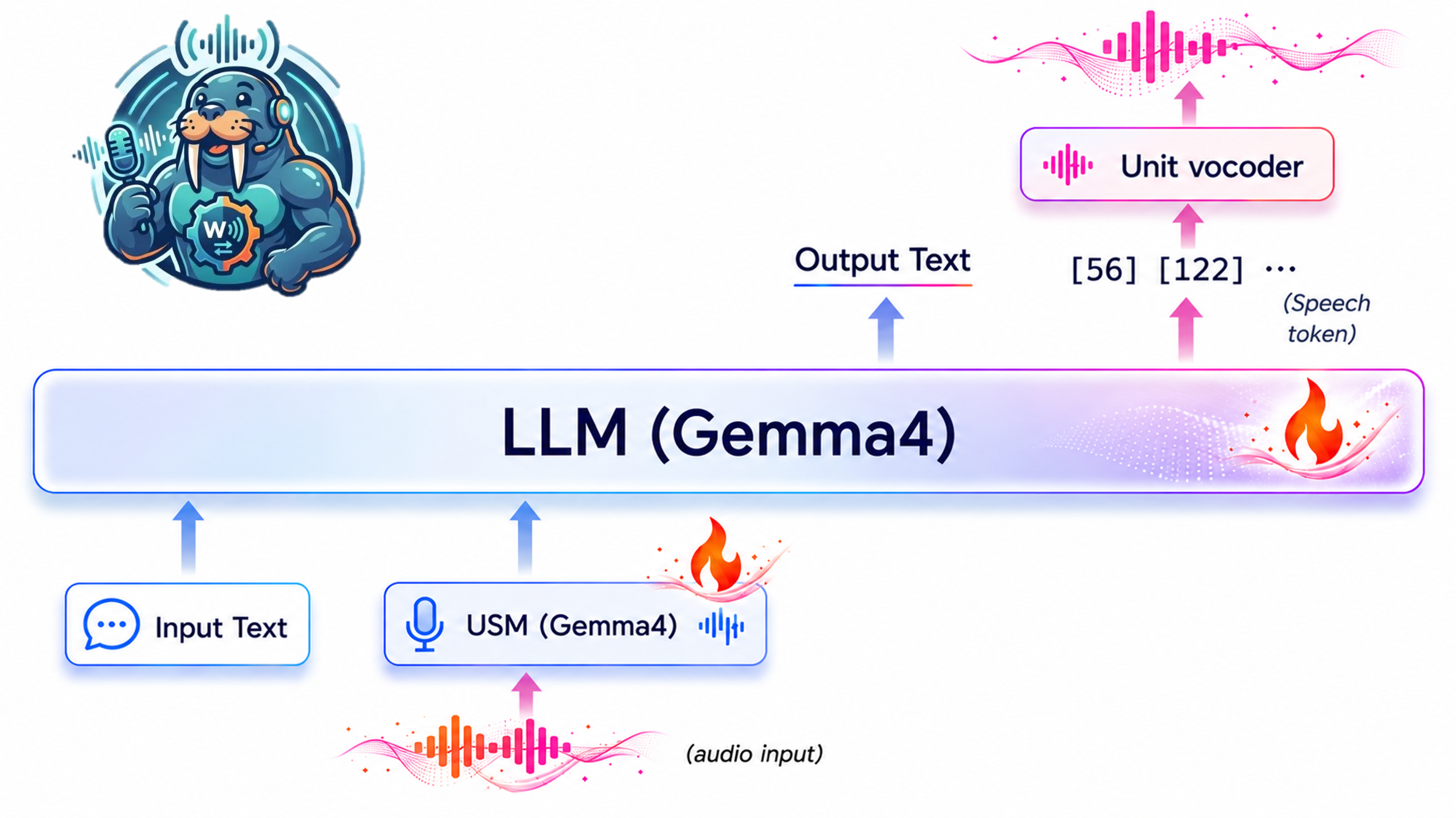}
  \caption{\textbf{Overview of Walrus, our most capable SpeechLLM yet.} Input speech is encoded by the USM encoder into continuous representations, which are passed to the decoder-only Gemma 4 backbone together with a natural-language task prompt. For recognition, translation, and understanding tasks, the backbone generates text. For TTS and S2ST, it generates discrete speech tokens at 25~Hz (325 bits per second), which a VQ decoder and a Vocos-based vocoder convert into a waveform. The vocoder is also conditioned on the source speech to transfer speaker identity and prosody.}
  \label{fig:architecture}
  \vspace{-4mm}
\end{figure}

\section{Introduction}
\label{sec:intro}
Speech is a natural interface between people and machines, and real applications rarely need only one skill.
A meeting assistant must transcribe overlapping speakers and keep track of who said what; a dubbing system must translate speech into another language and speak it back in a natural voice; a media platform must transcribe noisy entertainment audio and describe its music and sound effects.
Across these settings, systems also need to recognize emotion, answer questions about audio, and handle speakers who switch languages mid-sentence~\cite{ji2024wavchat, arch, sllm_survey, audiobench}.
Training and serving a separate model for each task is costly, and the resulting pipelines pass errors from one stage to the next; a single model that covers all of these tasks is cheaper to deploy and simpler to maintain.

Speech Large Language Models (SpeechLLMs) are a promising path to such a model.
Recent work improves them by scaling the model or redesigning the architecture~\cite{phi4-mm, kimi_audio, ghosh2026audio, intelligence2025amazon, xu2025qwen2}.
Far less is reported about the data: how it is cleaned, mixed, and scheduled across tasks.
This gap matters because the hardest cases are data problems: code-switched speech, overlapping speakers, entertainment audio, and reasoning-focused audio questions are rare in public corpora.
Tasks also differ greatly in token budget, so a naive mixture lets long generative outputs dominate the loss and weakens core recognition.
We take a different position: we argue that careful data engineering is a primary driver of quality, and that it can deliver state-of-the-art results without any new architecture.
We demonstrate this by holding the architecture fixed and putting all design effort into the data.

We present \textbf{Walrus}, a unified speech-and-audio foundation model with about 8B parameters.
Walrus is initialized from the \texttt{Gemma} \texttt{4} \texttt{E4B} model\footnote{\label{fn-gemma4-e4b}\href{https://huggingface.co/google/gemma-4-E4B}{\texttt{huggingface.co/google/gemma-4-E4B}}}~\cite{team2026gemma} and keeps the backbone architecture unchanged.
It supports automatic speech recognition (ASR), code-switching ASR (CS-ASR), multi-talker ASR (MT-ASR) with speaker diarization, machine translation (MT), automatic speech translation (AST), speech-to-speech translation (S2ST), text-to-speech (TTS), speech emotion recognition (SER), audio captioning, closed captioning, and general audio understanding.
For speech generation, Walrus predicts low-bitrate discrete speech tokens at only 325 bits per second; a lightweight token-to-waveform decoder converts these tokens into high-fidelity audio.

\begin{figure}
  \centering
  \includegraphics[width=0.75\linewidth]{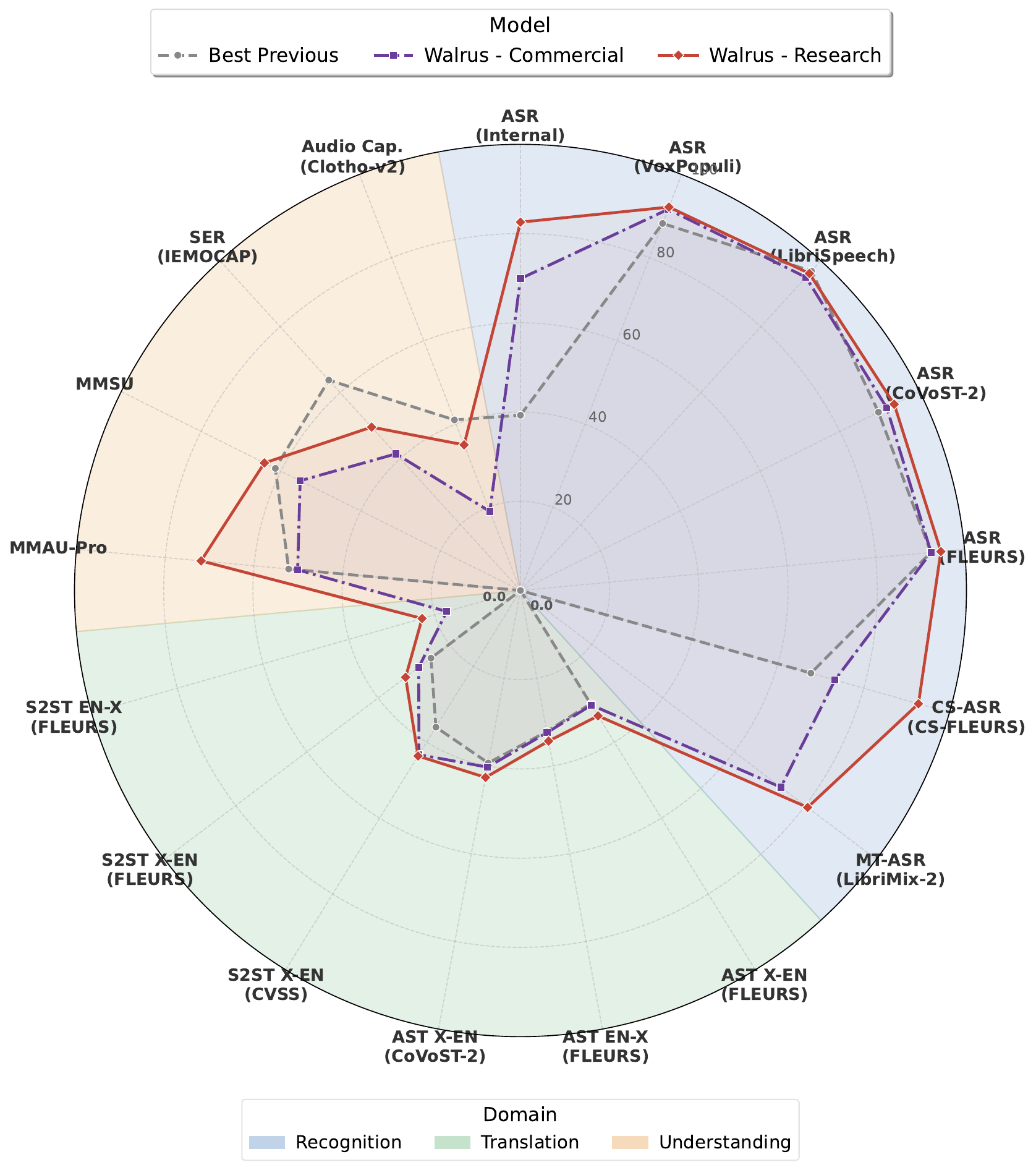}
  \caption{\textbf{Walrus against the best previous system (among AF3, Gemma4, Qwen2.5-Omni-7B, Phi4-MM) across recognition, translation, and understanding benchmarks.} Scores are normalized so that higher is better (ASR and CS-ASR shown as word/character accuracy, translation as BLEU or ASR-BLEU, understanding as accuracy, captioning as CLAP-Sim$\times$100). Both Walrus variants advance the frontier on most axes, and the Research variant extends it further, most notably on code-switching, internal entertainment ASR, and audio understanding.}
  \label{fig:radar}
  \vspace{-5mm}
\end{figure}

The core of our work is a three-stage data curriculum, where each stage adds new data and new tasks on top of a stable foundation.
We pair the curriculum with task-aware and dataset-aware weighting to balance tasks with very different token budgets and keep core skills strong as the model learns harder tasks.
From this curriculum, we build two variants: a \textit{Commercial-Ready} (\texttt{CR}) model trained only on commercially usable data, and a \textit{Research} (\texttt{R}) model trained on the full curriculum for the best overall quality.
The recipe leads to large and consistent gains over strong open SpeechLLMs of comparable size.
Figure~\ref{fig:radar} summarizes performance across recognition, translation, and understanding.
Both Walrus variants push the frontier beyond the best previous system on most axes.

The main results are as follows.

\textbf{Multilingual ASR.}
On VoxPopuli~\cite{wang2021voxpopuli},\footnote{Scores are reported as average across 5 major European languages: English (\texttt{EN}), French (\texttt{FR}), German (\texttt{DE}), Italian (\texttt{IT}), Spanish (\texttt{ES}).} Walrus reaches 92.2\% word accuracy (7.8 WER), against 88.2\% for the strongest comparable system, Gemma~4 (11.8 WER).
On FLEURS~\cite{conneau2023fleurs},\footnote{Scores are reported as average across Arabic (\texttt{AR}), Chinese (\texttt{ZH}), English (\texttt{EN}), French (\texttt{FR}), German (\texttt{DE}), Hindi (\texttt{HI}), Italian (\texttt{IT}), Japanese (\texttt{JA}), Korean (\texttt{KO}), Portuguese (\texttt{PT}), Spanish (\texttt{ES}).} it reaches 94.7\% word accuracy, ahead of 93.8\% for the $4\times$ larger Qwen3-Omni-30B~\cite{xu2025qwen3} and 92.0\% for Whisper Large-v3~\cite{radford2023robust}.

\textbf{Code-switching ASR.}
On CS-FLEURS~\cite{yan25c_interspeech} read speech, Walrus reaches 92.8\% character accuracy, far ahead of 67.7\% for Gemma 4 (25 points, 32.30 vs. 7.23 CER) and 61.3\% for Qwen3-Omni.

\textbf{Multi-talker ASR.}
Comparable open SpeechLLMs cannot transcribe overlapping speakers and produce no valid output.
On LibriMix clean 2-mix~\cite{cosentino2020librimix}, Walrus reaches 19.32 cpWER, half of the 37.04 for Qwen3-Omni-30B and far below the 59.34 of the recent dedicated VibeVoice-ASR~\cite{peng2026vibevoice}.
On Japanese Daily Talk~\cite{cejc}, it reaches 56.29 cpCER, against 71.99 for VibeVoice-ASR.
    
\textbf{Speech translation.}
On CVSS-C X$\to$En~\cite{jia2022cvss}, Walrus surpasses the much larger 30B Qwen3-Omni in S2ST (43.63 vs. 41.27 ASR-BLEU), while achieving very close performance on FLEURS X$\to$En and En$\to$X~\cite{conneau2023fleurs} (32.28/22.94 vs. 33.60/25.67, respectively). It achieves, however, better speaker similarity, prosody and emotion preservation on FLEURS X$\leftrightarrow$En compared to both larger Qwen3-Omni-30B and the dedicated SeamlessM4T v2~\cite{barrault2023seamless} (Table~\ref{tab:utmos-ssim-apcp}).

\textbf{Audio understanding.}
On MMSU~\cite{wang2026mmsu}, Walrus reaches 64.1\% accuracy, above 61.4\% for Audio-Flamingo-3~\cite{ghosh2026audio} and 60.6\% for Qwen2.5-Omni~\cite{xu2025qwen2}.
On MMAU-Pro~\cite{kumar2026mmau}, it reaches 71.92\%, above closed-sourced models like Gemini-Flash-3.7~\cite{team2023gemini} (70.40\%) and GPT-4o-Audio~\cite{hurst2024gpt} (52.50\%).

Moving from the Commercial-Ready to the Research variant improves every benchmark in Table~\ref{tab:master}.
For example, it lowers the code-switching character error rate by 19 points and the internal entertainment WER by 9 points.

Our main contributions are as follows.
First, we present a data-centric recipe for training a multi-task speech and audio model on a fixed backbone.
Second, we design a three-stage curriculum with task-aware and dataset-aware weighting, which grows from broad multilingual coverage to complex and domain-specific data.
Third, we describe the data pipelines behind the hardest tasks: synthetic code-switching, multi-talker mixtures, entertainment speech, conversational context, closed captioning, and audio reasoning.
Finally, we show through extensive evaluation that this recipe reaches state-of-the-art results on most benchmarks, and we provide a commercial-ready variant trained only on commercially usable data.

\section{Model}
\label{sec:model}
The Walrus architecture is deliberately simple; our contribution is the data strategy, not the network design.
This section describes the backbone (Sec.~\secref{sec:architecture}, including the audio representation used for input and output and the token-to-waveform decoder, and the supported tasks (Sec.~\secref{sec:tasks}).
Figure~\ref{fig:architecture} shows how the components connect.

\subsection{Architecture}
\label{sec:architecture}

\vspace{2mm}
\textbf{Backbone.}
Walrus is initialized from \texttt{Gemma} \texttt{4} \texttt{E4B},\footref{fn-gemma4-e4b} a decoder-only Transformer with about 4B effective parameters ($\approx$ 8B in total, including embeddings).
We keep the architecture unchanged and fine-tune all parameters on our data curriculum, using scaled dot-product attention (SDPA) for efficient training and inference.
A single set of weights serves every task, and tasks are selected through natural-language prompts.

\vspace{2mm}
\textbf{Speech input.}
Walrus keeps the speech input design of \texttt{Gemma} \texttt{4} \texttt{E4B}; the Universal Speech Model (USM) encoder~\cite{zhang2023google} converts the input speech into continuous representations, which are then fed to the decoder together with the text prompt.
We unfreeze the encoder during fine-tuning, so that it adapts to our audio domains.

\vspace{2mm}
\textbf{Speech output.}
For TTS and S2ST, Walrus generates discrete speech tokens autoregressively; we extend the backbone vocabulary with 8{,}192 dedicated speech tokens.
Such tokens are then converted back into a waveform via a separate decoder, which we adopt from \cite{futami2026kraken}.
It consists of a single-layer vector quantization (VQ) tokenizer and a GAN-based vocoder.
Decoupling token prediction from waveform synthesis keeps the backbone lightweight and lets the decoder specialize in high-fidelity generation and speaker transfer.

\vspace{0.5mm}
\textit{Speech VQ tokenizer.}
The tokenizer is a single-layer VQ module trained to reconstruct self-supervised speech features.
The target features come from the 20th layer of a pre-trained W2v-BERT 2.0 encoder~\cite{barrault2023seamless}, which empirically best captures semantic content.
A strided convolution reduces the feature rate from 50~Hz to 25~Hz; a three-layer Conformer~\cite{gulati2020conformer} encodes the features, and a single codebook with 8{,}192 entries in a 512-dimensional space quantizes them.
A symmetric Conformer decoder reconstructs the features.
We train the module end-to-end with a VQ-VAE objective~\cite{van2017neural}, which combines a cosine reconstruction loss and a commitment loss.
The resulting stream has 25 tokens per second at 13 bits each ($\log_2 8192 = 13$), which gives a bit-rate of 325 bits per second.

\vspace{0.5mm}
\textit{Vocoder.}
The GAN-based vocoder converts the VQ speech tokens into a waveform in a single non-autoregressive step.
It builds on UniCATS (\texttt{CTX-vec2wav})~\cite{du2024unicats} and replaces the generator with Vocos~\cite{siuzdak2023vocos} for higher synthesis quality.
The VQ tokens are first mapped back to continuous features by the VQ decoder, and then converted onto waveform through the Vocos decoder.
The vocoder is additionally conditioned on the source speech, using features from the 8th layer of W2v-BERT 2.0.
Lower layers of the encoder encode speaker identity more strongly, so this conditioning transfers the source speaker identity and prosody to the output.
It also relaxes the data requirement for expressive generation, because the model no longer needs source and target recordings with matched speaker and prosody.
We train the vocoder as in~\cite{futami2026kraken}, with multi-period and multi-band discriminators, a mel-spectrogram loss, a feature-matching loss~\cite{kong2020hifi, kumar2023high}, and a multi-resolution spectral loss~\cite{yamamoto2020parallel}.

\subsection{Supported tasks}
\label{sec:tasks}
Walrus handles a broad range of speech and audio tasks; Table~\ref{tab:tasks} groups them into four families.
Each task is specified through a natural-language prompt at both training and inference time.
During training, we draw the prompt for each sample from a pool of task-specific templates, which improves robustness to prompt wording; Appendix~\secref{app:prompts} lists the most frequently used prompt for each task.

The task families span the full pipeline from recognition to generation.

\vspace{2mm}
\textbf{Recognition.}
This family covers standard ASR together with all its conditioned variants.
Context-conditioned ASR receives preceding text as discourse context, while keyword-conditioned ASR receives a list of candidate keywords to bias transcription toward salient entities, including names and domain terms.
Timestamp-aware ASR predicts word or segment times, and a timestamp QA variant answers what is said within a given time span.
The family also includes code-switching ASR (CS-ASR), which handles utterances that mix two languages within a single sentence and transcribes each word in the phonetics of its own language.
Finally, it covers multi-talker ASR (MT-ASR) for audio with up to 4 speakers: overlapping-speech ASR that produces a speaker-attributed transcription, target-speaker ASR that transcribes only a chosen speaker, and speaker diarization.
We also support chain-of-thought variants, which first estimate the number of speakers and then transcribe each one, together with timestamped multi-talker transcription.

\vspace{2mm}
\textbf{Translation.}
This family covers text-to-text machine translation (MT), speech-to-text translation (AST), and speech-to-speech translation (S2ST).
Both AST and S2ST are supported in direct and cascaded forms: the direct form maps source speech to the target in one step, while the cascaded form first transcribes and then translates (and eventually synthesizes).
Context- and keyword-conditioned AST and S2ST variants carry the same discourse and entity hints as their recognition counterparts.

\vspace{2mm}
\textbf{Generation.}
This family covers text-to-speech (TTS).
Together with S2ST, generative tasks emit the low-bitrate discrete speech tokens described in Sec.~\secref{sec:architecture}, which the decoder then renders into audio.

\vspace{2mm}
\textbf{Understanding.}
This family covers audio captioning, closed captioning (CC), audio question answering, audio reasoning, and speech emotion recognition (SER).
Closed captioning spans several granularities, from a simple list of speech/non-speech events up to a full structured scene summary with speakers, sound effects, music, and timing (similar to~\cite{kumar2026tac}).
Audio reasoning covers five skill families, spanning temporal, speech-in-sound, speaker, spatial, and instruction-following questions (Sec.~\secref{subsec:understanding_supervision}).

\begin{table}
    \centering
    \caption{\textbf{Task families supported by Walrus, with language coverage.}
    ASR and its variants support more than 100 languages.
    CS-ASR supports English-Japanese, and English or Japanese mixed with Chinese, French, German, Hindi, Italian, or Spanish (both as major and minor script).
    MT-ASR supports 2-, 3-, and 4-speaker mixtures in Chinese, English, French, German, Italian, Japanese, and Spanish.
    AST and S2ST support translation from 10 languages into English or Japanese, and from English or Japanese into the same 10 languages.
    Closed captioning supports Chinese, English, French, German, Italian, Japanese, and Spanish.}
    \label{tab:tasks}
    \vspace{1mm}
    \begin{tabular}{ll}
        \toprule
        \textbf{Task family} & \textbf{Supported tasks} \\
        \midrule
        \rowcolor{stdblue}
        \textbf{Recognition}   & ASR, Context-ASR, Keyword-ASR, Timestamp-ASR, \\
        \rowcolor{stdblue}
                      & CS-ASR, MT-ASR, Target-Speaker ASR, Diarization \\
        \rowcolor{specgreen}
        \textbf{Translation}   & MT, AST (direct, cascaded), S2ST (direct, cascaded) \\
        \rowcolor{envred}
        \textbf{Generation}    & TTS \\
        \rowcolor{accentyellow}
        \textbf{Understanding} & Audio Captioning, Closed Captioning, Audio QA, Timestamp-QA, \\
        \rowcolor{accentyellow}
                      & Audio Reasoning, Instruction Following, SER \\
        \bottomrule
    \end{tabular}
    \vspace{-3mm}
\end{table}

\section{Data curriculum}
\label{sec:curriculum}
The data curriculum is the core of Walrus; we hold the architecture fixed, so that gains can be traced to the data.
This section describes how we compose, weight, and stage the training data across all four task families.
We first state our design principles and the resulting task distribution (Sec.~\secref{sec:principles}-\secref{sec:distribution}).
We then describe the shared training setup (Sec.~\secref{sec:common-setup}), the three stages (Sec.~\secref{sec:stage1} to Sec.~\secref{sec:stage3}), and the two trained variants (Sec.~\secref{sec:variants}).
Finally, we describe the data-preparation pipelines (Sec.~\secref{sec:data-prep}).

\subsection{Design principles}
\label{sec:principles}
Our data strategy follows three principles that apply to every task family.

\vspace{2mm}
\textbf{Grow from broad and clean to complex and specialized.}
Early training uses well-known, high-quality corpora that teach stable core skills while retaining Gemma 4's foundation capabilities.
Later training adds harder data, such as overlapping speech, code-switching, and in-domain entertainment synthetic audio.
This order lets the model learn easy skills before it faces difficult ones.

\vspace{2mm}
\textbf{Balance the mix with explicit weights.}
Tasks differ greatly in the number of tokens they produce: generative tasks such as TTS and S2ST emit long token sequences, so they can dominate the training loss; recognition tasks emit far fewer tokens, yet they form the backbone of the model.
We therefore give generative tasks low weights in the early stages, and raise them only after the recognition core is stable.

\vspace{2mm}
\textbf{Protect core skills at every stage.}
We keep foundation data in all three stages, and ASR keeps the highest task weight in every stage.
As a result, adding new tasks in later stages does not erode the recognition skills built in the first stage.

\begin{figure}
  \centering
  \begin{subfigure}[t]{0.50\linewidth}
    \centering
    \includegraphics[width=\linewidth]{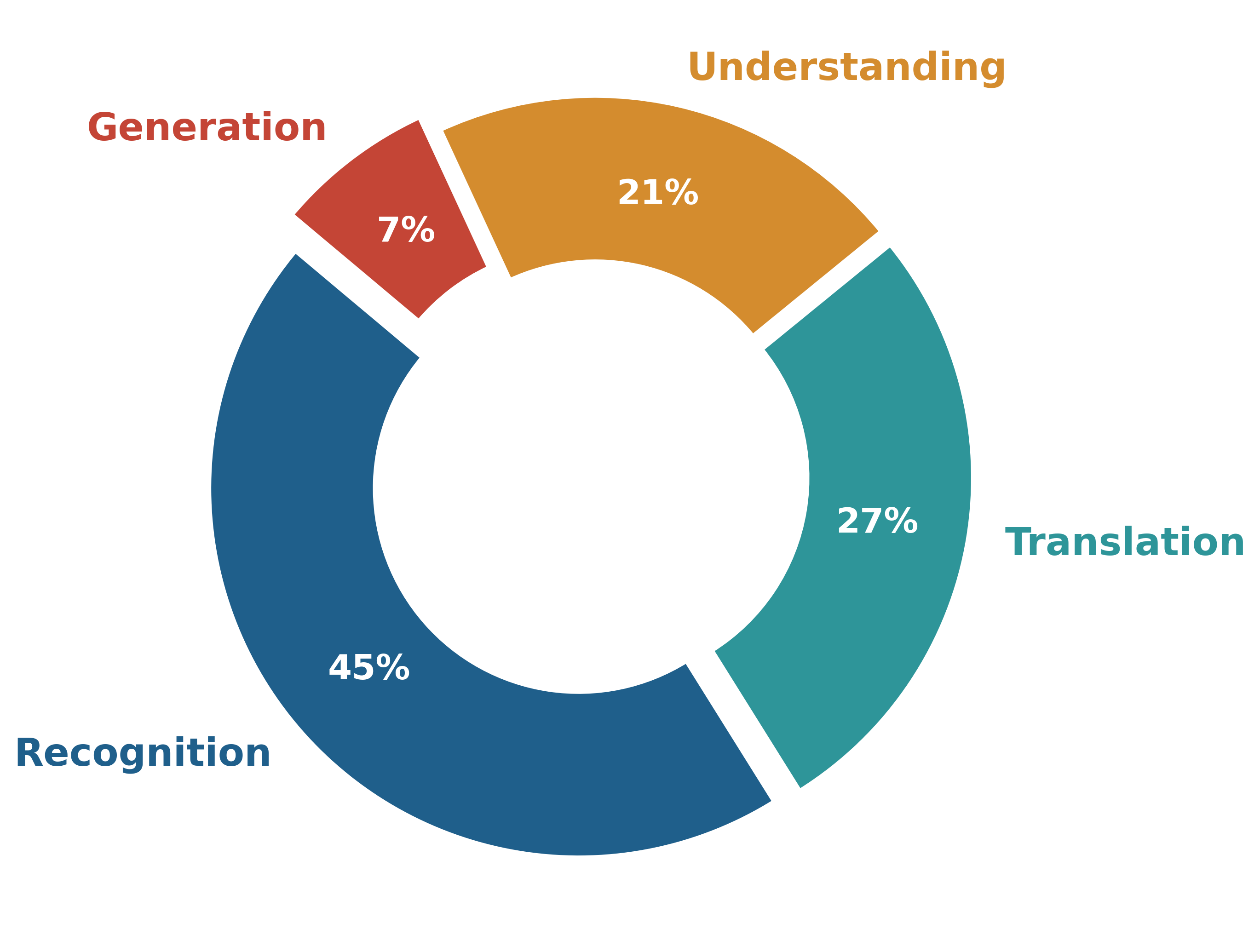}
    \caption{Overall task distribution across the four task families.}
    \label{fig:dist-overall}
  \end{subfigure}
  \hfill
  \begin{subfigure}[t]{0.98\linewidth}
    \centering
    \includegraphics[width=\linewidth]{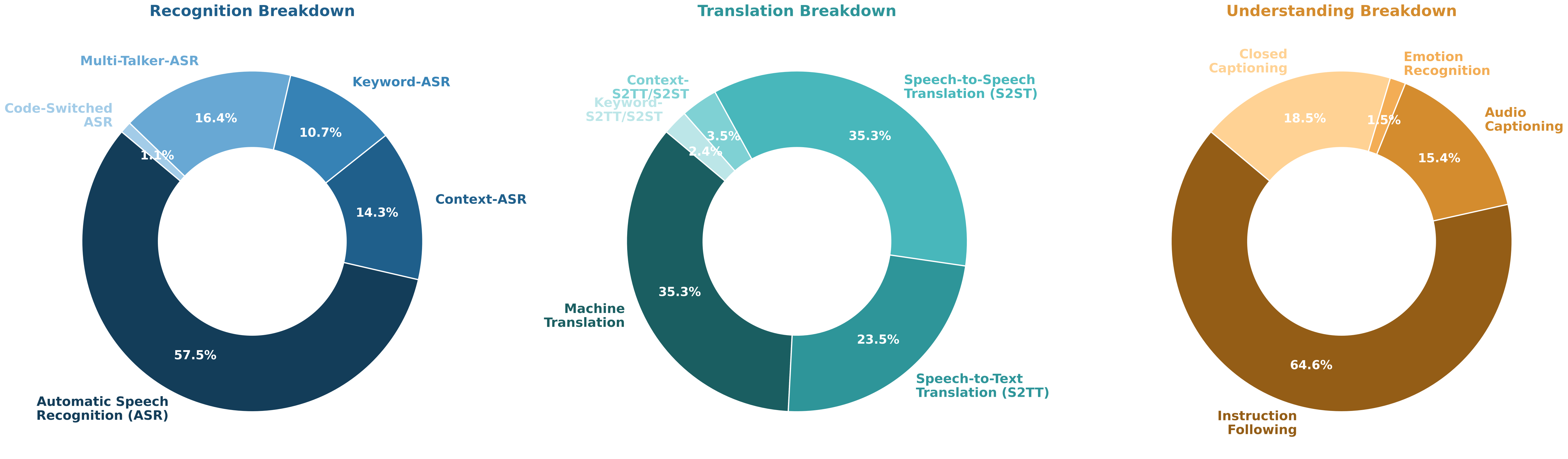}
    \caption{Detailed breakdown within the recognition, translation, and understanding families.}
    \label{fig:dist-detail}
  \end{subfigure}
  \caption{\textbf{Training data composition of Walrus.} Recognition dominates the mix (45\%), followed by translation (27\%) and understanding (21\%), while generation is deliberately kept small (7\%). The detailed breakdown shows that standard ASR and instruction-following data anchor the recognition and understanding families. This distribution reflects our data-centric design: core recognition is heavily represented to protect it from forgetting, while token-heavy generative tasks are downweighted.}
  \label{fig:data-distribution}
  \vspace{-3mm}
\end{figure}

\subsection{Task distribution}
\label{sec:distribution}
Figure~\ref{fig:dist-overall} shows the overall task distribution of the training data across the four families.
Recognition is the largest family with 45\%, followed by translation with 27\%, understanding with 21\%, and generation with 7\%.
This mix reflects our priorities: a dominant recognition core, strong translation and understanding coverage, and a smaller but sufficient share of generation.

\begin{table}[!ht]
    \centering
    \caption{\textbf{Training schedule per stage.} Each stage is initialized from the previous checkpoint. Warmup and decay are given as a fraction of total steps.}
    \label{tab:stages}
    \vspace{1mm}
    \begin{tabular}{lcccc}
        \toprule
        \textbf{Stage} & \textbf{Steps} & \textbf{Learning rate} & \textbf{Warmup} & \textbf{Decay} \\
        \midrule
        1: Foundation      & 87{,}500  & $2\times10^{-6}$ & 1\% & 5\% \\
        2: Expansion       & 135{,}000 & $2\times10^{-6}$ & 1\% & 5\% \\
        3: Specialization  & 150{,}000 & $1\times10^{-6}$ & 1\% & 5\% \\
        \bottomrule
    \end{tabular}
    \vspace{-3mm}
\end{table}

The families are themselves diverse (Figure~\ref{fig:dist-detail}); all percentages below are shares of the respective family.
In recognition, standard ASR makes up 57.5\% of the total amount, with the remainder split across multi-talker ASR (16.4\%), context ASR (14.3\%), keyword ASR (10.7\%), and code-switching ASR (1.1\%).
In translation, MT and S2ST each contribute 35.3\%, AST accounts for 23.5\%, and context- and keyword-conditioned variants make up the rest.
In understanding, instruction following, audio QA, and reasoning represent 64.6\%, followed by closed captioning (18.5\%), audio captioning (15.4\%), and emotion recognition (1.5\%).

These breakdowns show that the mix is broad within each family, not concentrated in a single task.

\subsection{Common setup}
\label{sec:common-setup}
All three stages share the same optimization setup.
We use the AdamW optimizer with $\beta_1 = 0.9$, $\beta_2 = 0.95$, $\epsilon = 10^{-8}$, and a weight decay of 0.1.
We clip the gradient norm to 1.0 and use scaled dot-product attention (SDPA) for efficient training and inference.
We stream the data from WebDataset shards.
We use a per-device batch size of 8 with 2 gradient accumulation steps, and a maximum sequence length of {1},024 tokens; on 128 GPUs, this gives an effective batch size of 2{,}048 samples.
Table~\ref{tab:stages} lists the learning rate and schedule of each stage.

We shape the data mix with two independent controls: task weights scale the contribution of each task to the loss, whereas dataset weights scale how often each dataset is sampled, so a weight of 0.5 halves the effective size of a corpus in the mixture.
Together, the two controls let us set the effective data distribution independently of the raw corpus sizes.
Table~\ref{tab:weights} summarizes the task weights of the three stages.

\subsection{Stage 1: Foundation}
\label{sec:stage1}
The first stage builds a broad multilingual foundation across all four task families.
It covers ASR, MT, AST, S2ST, TTS, SER, audio captioning, and audio QA, in many languages.
It does not yet include timestamp ASR, code-switching ASR, multi-talker ASR, diarization, or closed captioning.

We draw on well-established public corpora and internal data.
For speech recognition and translation, these comprehend CommonVoice (\texttt{CV14})~\cite{ardila2020common}, VoxPopuli~\cite{wang2021voxpopuli}, LibriSpeech~\cite{librispeech} and Multilingual LibriSpeech~\cite{pratap2020mls}, FLEURS~\cite{conneau2023fleurs}, CoVoST-2~\cite{wang2020covost}, AISHELL-1~\cite{bu2017aishell}, CSJ~\cite{maekawa03_sspr}, YODAS-Granary~\cite{raokoluguri25_interspeech}, and IndicVoices-ST~\cite{javed2024indicvoices}.
For TTS, we use LibriTTS~\cite{zen19_interspeech}, VCTK~\cite{cvtk}, and AISHELL-3~\cite{shi2020aishell}.
We further leverage internal datasets for audio captioning, and a small subset of SIFT-50M~\cite{pandey2025sift} for instruction following.
We clean the transcripts as described in Sec.~\secref{sec:data-prep}.

The task weights make recognition the priority (Table~\ref{tab:weights}).
ASR has the highest weight (5.0), followed by context and keyword ASR (3.0).
AST has a weight of 2.0, and MT and S2ST have a weight of 0.5 each.
TTS has a weight of only 0.25, so that its long token sequences do not dominate the loss.
Audio QA, audio captioning, and SER have a weight of 1.0 each.
All datasets are sampled at their natural size, and we do not yet apply dataset weights.
This stage teaches stable core skills before we introduce harder tasks.

\begin{table}[!ht]
    \centering
    \caption{\textbf{Task weights across the three stages.} Weights scale each task's contribution to the loss. A dash means the task is not yet present. Recognition is upweighted throughout to protect the core, while generation and understanding are raised in Stage 3.}
    \label{tab:weights}
    \vspace{1mm}
    \begin{tabular}{lccc}
        \toprule
        \textbf{Task} & \textbf{Stage 1} & \textbf{Stage 2} & \textbf{Stage 3} \\
        \midrule
        \rowcolor{stdblue}
        \multicolumn{4}{@{}c}{\textit{Recognition}} \\
        ASR                    & 5.0 & 5.0 & 4.0 \\
        Context / Keyword ASR  & 3.0 & 3.0 & 3.0 \\
        Timestamp ASR          & --  & 1.0 & 1.0 \\
        CS-ASR                 & --  & 1.0 & 1.0 \\
        MT-ASR                 & --  & 1.0 & 1.0 \\
        Diarization            & --  & 1.0 & 1.0 \\
        \midrule
        \rowcolor{specgreen}
        \multicolumn{4}{@{}c}{\textit{Translation}} \\
        AST (and variants)     & 2.0 & 2.0 & 2.0 \\
        MT                     & 0.5 & 0.5 & 0.5 \\
        S2ST (and variants)    & 0.5 & 0.5 & 2.0 \\
        \midrule
        \rowcolor{envred}
        \multicolumn{4}{@{}c}{\textit{Generation}} \\
        TTS                    & 0.25 & 0.25 & 2.0 \\
        \midrule
        \rowcolor{accentyellow}
        \multicolumn{4}{@{}c}{\textit{Understanding}} \\
        Audio QA / Reasoning   & 1.0 & 1.0 & 1.0 \\
        Audio captioning       & 1.0 & 1.0 & 2.0 \\
        Closed captioning      & --  & 1.0 & 2.0 \\
        SER                    & 1.0 & 1.0 & 2.0 \\
        \bottomrule
    \end{tabular}
    \vspace{-3mm}
\end{table}

\subsection{Stage 2: Expansion}
\label{sec:stage2}
The second stage starts from the Stage-1 checkpoint and adds new tasks to every family.
The foundation data stays in the mix with reduced dataset weights, so that the new data has room in the mixture.

In recognition, we add multi-talker and target-speaker ASR.
The data comes from real meetings (AMI~\cite{kraaij2005ami}, AliMeeting~\cite{alimeeting}, and AISHELL-4~\cite{fu2021aishell}) and from simulated mixtures (LibriMix~\cite{cosentino2020librimix}).
We also add timestamp prediction, diarization, and synthetic code-switching ASR.
For understanding, we add synthetic closed captioning and richer audio QA, including additional SIFT-50M samples~\cite{pandey2025sift} and MECAT~\cite{niu2025mecat}.
We also add the UltraVoice~\cite{tu2025ultravoice} and MSP-Podcast~\cite{busso2026msp} corpora, which provide instruction-style speech and emotion data.
Across recognition and translation, we add dialogue-level context tasks built from synthetic dialogues (Sec.~\secref{sec:conversational-supervision}).
This stage teaches the model to handle overlapping speakers, mixed languages, timing, and conversational context.

We also introduce dataset weights in this phase.
They let us lower the influence of large or redundant corpora, so that these corpora do not crowd out smaller, high-value data.
For example, we halve the weight of very large ASR corpora such as Common Voice and YODAS-Granary, while giving the newly introduced tasks full weight.
The task weights of the shared tasks stay the same as in Stage 1, and the new tasks get a task weight of 1.0;
ASR therefore keeps the highest task weight.

\subsection{Stage 3: Specialization}
\label{sec:stage3}
The third stage starts from the Stage-2 checkpoint; it keeps the data of the first two stages and adds new data that closes specific gaps and covers target domains.
This also includes corpora that are not commercially usable.

In recognition, we add synthetic entertainment speech, low-resource languages, and more code-switching data.
We also add a large number of synthetic multi-talker mixtures with two to four speakers, covering seven languages: Chinese, English, French, German, Italian, Japanese, and Spanish.
In translation, we add more S2ST data: SpeechMatrix~\cite{duquenne2023speechmatrix}, Europarl-ST~\cite{iranzo2020europarl}, and internal dubbing-aligned data built with SpeechVecAlign~\cite{meng2025speech}.
For closed captioning, we add the remaining segment libraries (Sec.~\secref{subsec:understanding_supervision}).

We halve the learning rate to $1\times10^{-6}$, because a model with strong core skills needs refinement rather than large updates.
We also rebalance the task weights to strengthen generation and understanding: we lower the ASR weight from 5.0 to 4.0, while raising the weights of TTS (0.25 to 2.0), S2ST (0.5 to 2.0), audio captioning, closed captioning, and SER (each 1.0 to 2.0).
The ASR weight stays the highest, so recognition remains the core of the model.
Sec.~\secref{sec:data-prep} describes the data-preparation and filtering pipelines for these corpora.

\subsection{Training scale and variants}
\label{sec:variants}
Walrus is trained on over 600M samples that cover more than 20 task variants and 100 languages.
The run has three consecutive stages; each one is initialized from the previous checkpoint and uses a different random seed, which changes the data shuffling.
Training takes about 30 days on 128 NVIDIA H100 GPUs, which is about 92{,}000 GPU hours.

We derive two variants from the curriculum.
The \textit{Commercial-Ready} (\texttt{CR}) variant is the Stage-2 checkpoint, trained only on commercially usable data, making it suitable for deployment.
The \textit{Research} (\texttt{R}) variant is the Stage-3 final checkpoint; it uses all data, including non-commercial corpora, and delivers the best overall quality, achieving state-of-the-art performance.

The two variants share the same backbone, but differ in training stage, data license, and task weights.
The \textit{research} variant improves on the \textit{commercial-ready} variant on every benchmark in Table~\ref{tab:master}.
The largest gains are in CS-ASR (about 19 points lower CER on CS-FLEURS read speech), internal entertainment ASR (about 9 points lower average WER across 6 languages), and MMAU-Pro (more than 20 points higher accuracy).
These gains come from Stage 3, which adds high-quality data while rebalancing the task weights.

\subsection{Data preparation}
\label{sec:data-prep}
Raw corpora are rarely usable as they are, and public data is missing for several of our tasks.
We therefore clean the source data and build task-specific supervision for every task family; where suitable data does not exist at scale, we generate it with dedicated pipelines.
This applies to code-switching, multi-talker conversation, context- and keyword-conditioned recognition and translation, closed captioning, and reasoning-focused audio QA.

The following subsections describe these pipelines.
Sec.~\secref{sec:transcript-cleaning} describes transcript cleaning, which applies to all speech corpora.
Sec.~\secref{sec:recognition-supervision} covers recognition, Sec.~\secref{sec:conversational-supervision} summarizes synthetic dialogues, Sec.~\secref{sec:translation-supervision} covers translation, and Sec.~\secref{subsec:understanding_supervision} highlights understanding.
 Appendix~\secref{app:prompts} lists example task prompts, and Appendix~\secref{app:data-summary} summarizes all pipelines in one table.

\subsubsection{Transcript cleaning}
\label{sec:transcript-cleaning}
We apply two steps to the transcripts of our speech corpora: restoration of casing and punctuation, and text normalization.
Clean and readable text benefits every task that uses these transcripts.

\paragraph{Casing and punctuation restoration.}
Many ASR corpora provide only lowercase transcripts without punctuation.
A model trained on such text learns to drop casing and punctuation, which hurts readability and downstream tasks.
We restore both with a large language model, \texttt{Qwen3.5-9B},\footnote{\label{fn-qwen35-9b}\href{https://huggingface.co/Qwen/Qwen3.5-9B}{\texttt{huggingface.co/Qwen/Qwen3.5-9B}}} used as a text normalizer.
For each transcript, we build a few-shot prompt with language-specific examples.
The prompt has one strict rule: add casing and punctuation only, and never add, remove, or change a word.
We run the restoration separately for each language, and we skip transcripts that already contain mixed case.

A language model can silently rewrite words, so we verify every output: we remove casing and punctuation from the original and the restored transcript, and we compute the character error rate between them.
We keep a restored transcript only if this CER is at most 0.05: this limits the edits to the surface form and filters out outputs where the model changed the content.
We also discard outputs whose length differs too much from the original, as a second guard against hallucination.
The filter retains the large majority of samples (90-95\% for most datasets) while removing the few cases where the normalizer altered content.

\paragraph{Text normalization.}
After restoration, we apply a normalization pipeline with separate rules for three language groups: European languages, CJK languages (Chinese, Japanese, and Korean), and others.
These scripts differ in spacing, punctuation, and digit conventions, so each group needs its own rules; we tuned the rules by manually inspecting samples from each group.
The pipeline gives consistent and readable references across the multilingual mix.

\subsubsection{Recognition supervision}
\label{sec:recognition-supervision}
This subsection describes the data for the recognition tasks that need dedicated pipelines.
Context ASR and keyword ASR are built from cleaned ASR corpora.
Code-switching ASR is fully synthetic, and multi-talker ASR combines real and simulated recordings.
Entertainment speech combines curated available data with synthetic augmentation.
Sec.~\secref{sec:conversational-supervision} adds more context- and keyword-conditioned data from synthetic dialogues.

\paragraph{Context ASR.}
Standard ASR treats each utterance on its own, but real speech depends on what was said before.
We build context ASR data from the document structure of read-speech corpora.
We group utterances by chapter (for MLS~\cite{pratap2020mls}) or by domain (for internal datasets), and sort them in reading order.
For each target utterance, we randomly attach between one and five preceding utterances as text context.
The model is then asked to transcribe the target utterance given this context.
This teaches the model to use prior discourse, which helps with proper nouns, topic-specific vocabulary, and consistency across a sequence of utterances.

\paragraph{Keyword ASR.}
In keyword-biased ASR, the model receives a list of candidate keywords together with the audio.
We extract candidate keywords from transcripts using named-entity recognition, keeping people, places, organizations, and other proper nouns.
We normalize and de-duplicate the entities, and keep only those that appear in at least two utterances, which removes rare noise.
For each utterance, the keyword list contains the entities that the utterance actually includes (positives) and randomly sampled entities that it does not contain (distractors).
We draw the list size from $\{0, 10, 20, 50, 100\}$, so that the model sees short and long keyword lists, as well as no list at all.
We further shuffle each list, so that the position of a keyword carries no information.
This teaches the model to use helpful keywords and to ignore irrelevant ones, which matters for domain-specific speech.

\paragraph{Code-switching ASR.}
Code-switched speech, where a speaker mixes two languages in one utterance, is rare in public corpora.
We therefore generate both the text and the audio in three steps: code-switched text generation, segment-based speech synthesis, and quality filtering.
We cover 13 language pairs, with a focus on English and Japanese mixed with European and Asian languages.

\textit{Text generation.}
We generate the sentences with a large language model, \texttt{Gemma} \texttt{4} \texttt{26B}.\footnote{\label{fn-gemma4-26b}\href{https://huggingface.co/google/gemma-4-26B-A4B-it}{\texttt{huggingface.co/google/gemma-4-26B-A4B-it}}}
For each language pair, we prompt the model to write one natural sentence that is mostly in a matrix language with short embedded phrases from the other language.
The prompt enforces realistic switching: it allows switches only where a bilingual speaker would use them, such as technical terms, loanwords, or emotional expressions, and limits each switch to a few words.
We guide the model with hand-written few-shot examples for each language pair and each switching direction.
To increase diversity, we vary the topic across 25 everyday domains, such as work, travel, food, and technology.
After generation, we remove duplicate sentences so that each text is unique.

\textit{Speech synthesis.}
A single-language TTS model cannot pronounce a mixed sentence reliably, because each word must follow the phonetics of its own language; we therefore synthesize each sentence in monolingual segments.\footnote{We employ Qwen3-TTS (\href{https://huggingface.co/Qwen/Qwen3-TTS-12Hz-1.7B-Base}{\texttt{huggingface.co/Qwen/Qwen3-TTS-12Hz-1.7B-Base}}) and a fast in-house variant of Chatterbox-TTS (\href{https://github.com/resemble-ai/chatterbox}{\texttt{github.com/resemble-ai/chatterbox}}). We decide which to use depending on the language pair, following preliminary results from a pilot study.}
First, we detect which words belong to the embedded language, using a per-language word list built from the generated data.
Next, we group consecutive words of the same language into segments, and we merge very short segments into their neighbors to avoid irregular, discontinuous audio.
Then, we synthesize each segment with the correct language tag, and we join the segments with a random short silence gap.
Every word in the resulting audio carries the phonetics of its own language.

\textit{Speaker control.}
For each utterance, we sample one reference speaker from a pool of high-quality FLEURS speakers of the matrix language.
We select the pool with the \texttt{NISQA} quality score~\cite{mittag21_interspeech}, and we keep only clean, high-fidelity recordings.
We clone the reference voice for all segments of the utterance, so that the whole utterance has one consistent speaker.
Sampling a new speaker for each utterance gives broad speaker variety across the dataset.

\textit{Quality filtering.}
We apply two filters to the synthesized speech.
First, we score the naturalness of each utterance with \texttt{NISQA}, and we discard utterances below a 3.50 threshold.
Second, we check the intelligibility; we transcribe each utterance with \texttt{Whisper} \texttt{Large-v3} \texttt{Turbo}\footnote{\label{fn-whisper}\href{https://huggingface.co/openai/whisper-large-v3-turbo}{\texttt{huggingface.co/openai/whisper-large-v3-turbo}}}~\cite{radford2023robust} and compute the WER against the intended text (or CER, depending on the language).
We discard utterances whose error rate is above a 5\% threshold.
The remaining audio and its known code-switched text form the training pairs.

\paragraph{Multi-talker ASR.}
Real-world conversations and media content involve multiple speakers with overlapping speech.
We build data for four related tasks: overlapping-speech ASR with speaker-attributed output, target-speaker ASR, speaker diarization, and timestamped variants of these tasks.
With this data, the model learns to transcribe the content, to separate the speakers, and to predict when each speaker talks.
All clips are at most 30 seconds long, which is the input limit of Walrus.

We combine real and simulated conversations.
The real recordings come from AMI~\cite{kraaij2005ami}, AliMeeting~\cite{alimeeting}, and AISHELL-4~\cite{fu2021aishell}.
We split the full recordings into 30-second clips with the DiCoW recipe\footnote{\href{https://github.com/BUTSpeechFIT/mt-asr-data-prep.git}{\texttt{github.com/BUTSpeechFIT/mt-asr-data-prep.git}}}~\cite{dicow}.
For simulated conversations with overlapped speech, we rely on public recipes and further scale the data with FastMSS~\cite{fastmss}.
We first adopt the public LibriMix~\cite{cosentino2020librimix} and LibriSpeechMix~\cite{kanda2020serialized} corpora, then use FastMSS to cut single-speaker utterances at the word level and to rearrange them into multilingual conversations with two, three, or four speakers, including overlap and turn-taking.
The single-speaker sources are LibriSpeech for English, an internal corpus for Japanese, and MLS for French, Spanish, German, and Italian.
Because MLS and the internal Japanese corpus lack the word- and character-level alignments that FastMSS requires, we generate them with \texttt{Qwen3-ForcedAligner-0.6B}\footnote{\href{https://huggingface.co/Qwen/Qwen3-ForcedAligner-0.6B}{\texttt{huggingface.co/Qwen/Qwen3-ForcedAligner-0.6B}}}~\cite{qwen3-asr}.

\paragraph{Entertainment speech.}
Entertainment audio, such as anime, games, movies, and TV, is a key target domain, yet clean labeled data for it is scarce.
We build this data in two ways.
First, we curate in-domain speech from large, permissively licensed collections such as YODAS-Granary~\cite{raokoluguri25_interspeech}, and we pass it through the cleaning and filtering pipeline of Sec.~\secref{sec:transcript-cleaning}.
Second, we synthesize entertainment-style speech by augmenting existing corpora with domain-characteristic effects, such as high-pitched or stylized voices, expressive prosody, and non-speech vocal and background sounds.
This exposes the model to the acoustic conditions of entertainment content that read-speech corpora rarely contain.
We use the resulting data for ASR and, because the synthetic effects and background events are known by construction, also for the closed-captioning pipeline of Sec.~\secref{subsec:understanding_supervision}.

\subsubsection{Conversational supervision via synthetic dialogues}
\label{sec:conversational-supervision}
Real conversational data with reliable context, translations, and keyword annotations is scarce.
We therefore build synthetic multi-turn dialogues, following the framework of~\cite{koudounas2025deepdialogue}.
The dialogues provide context- and keyword-conditioned supervision for ASR, AST, and S2ST.
The pipeline has five steps: dialogue generation, dialogue-level filtering, speech synthesis, quality gating, and task construction.

\textit{Dialogue generation.}
We generate dialogues between two instruction-tuned LLMs across 40 everyday domains, such as travel, food, technology, and health.
An emotion-chain manager drives each dialogue, so that emotions evolve naturally from turn to turn and stay appropriate to the domain.
We use three model pairings to increase diversity: \texttt{Gemma} \texttt{4} \texttt{26B}\footref{fn-gemma4-26b}--\texttt{Gemma} \texttt{4} \texttt{26B}, \texttt{Qwen3.5} \texttt{9B}\footref{fn-qwen35-9b}--\texttt{Gemma} \texttt{4} \texttt{26B}, and \texttt{Qwen3.5} \texttt{9B}--\texttt{Qwen3.5} \texttt{9B}.
Each dialogue has two to five turns, because smaller models tend to hallucinate in longer dialogues~\cite{koudounas2025deepdialogue, laban2026llms}.
Each textual dialogue is then translated from English to 11 different languages using \texttt{Qwen3.5} \texttt{27B}.\footnote{\href{https://huggingface.co/Qwen/Qwen3.5-27B}{\texttt{huggingface.co/Qwen/Qwen3.5-27B}}}

\textit{Dialogue-level filtering.}
We filter the generated dialogues with an ensemble of LLM judges, as in~\cite{koudounas2025deepdialogue}.
Each judge gives a binary coherence decision and a five-point quality rating.
We keep a dialogue only if all judges find it coherent and give it the highest rating.

\textit{Speech synthesis.}
We synthesize each turn with the in-house \texttt{Chatterbox} variant described in Sec.~\secref{sec:recognition-supervision}.
Speaker voices are cloned from Common Voice (\texttt{CV14}) reference clips~\cite{ardila2020common}.
Each dialogue has two distinct speakers, balanced by age and gender; we use fallback strategies when the pool lacks a matching speaker, which keeps the speaker pool diverse.

\textit{Quality gating.}
We apply strict quality gates to every synthesized turn.
We transcribe the turn with \texttt{Whisper} \texttt{Large-v3} \texttt{Turbo}\footref{fn-whisper} and compute the WER or CER, depending on the language.
For translation tasks, we also score the target text with a reference-free quality estimator, which differs depending on the training stage.
In Stage~2, we use an LLM judge (\texttt{Gemma} \texttt{4} \texttt{26B}),\footref{fn-gemma4-26b} which we prompt to output a score on the same 0-1 scale as \texttt{COMET-Kiwi}~\cite{rei2022cometkiwi}.
In Stage~3, we use \texttt{COMET-Kiwi} directly.\footnote{\href{https://huggingface.co/Unbabel/wmt22-cometkiwi-da}{\texttt{huggingface.co/Unbabel/wmt22-cometkiwi-da}}}
We keep a turn only if its error rate is at most 0.1 and its translation quality score (\texttt{COMET-Kiwi} or its LLM equivalent) is at least 0.80.
These gates keep only clean and faithful turns.

\textit{Task construction.}
We build context- and keyword-conditioned tasks from the filtered turns.
For keyword tasks, we extract salient entities (people, locations, and organizations) with a multilingual named-entity recognizer, then present them together with distractor keywords in a list of ten to fifty items.
For context tasks, we instead prepend the previous turns of the dialogue.
This gives data for six tasks: context ASR/AST/S2ST, keyword ASR/AST/S2ST.
Translation tasks are generated in both direct and cascaded forms.
This gives the model rich supervision for handling conversational context and salient keywords, which standard corpora rarely provide.

\subsubsection{Translation supervision}
\label{sec:translation-supervision}
Beyond the synthetic dialogues, we build a large pool of parallel data for MT, AST, and S2ST.
We first mine and filter parallel text, then synthesize speech for the retained text pairs, which gives the supervision for AST and S2ST.

\paragraph{Parallel text mining.}
We mine parallel text from the OPUS collection~\cite{tiedemann2004opus}.
Because our target includes a commercial-ready model, we first restrict OPUS to datasets with clear, permissive licenses.\footnote{We focus on English- and Japanese-centric pairs, covering mainly Arabic, Chinese (including Traditional Chinese), French, German, Hindi, Italian, Korean, Portuguese, Russian, and Spanish. We use only the main variety of each language, which avoids tiny dialect-specific splits.}
We extract the source-target sentence pairs and shuffle them, which removes any corpus- or document-level ordering.
Mined parallel data is noisy, so we score every pair for cross-lingual semantic similarity.
We embed both sides with a multilingual e5 sentence encoder\footnote{\href{https://huggingface.co/intfloat/multilingual-e5-large}{\texttt{huggingface.co/intfloat/multilingual-e5-large}}} and compute the cosine similarity of the two embeddings.
For each language pair, we sort the pairs by this score and keep the top decile: this removes most misaligned, partial, and low-quality translations.
We also drop pairs whose source sentence starts with a non-letter character, because these are usually markup, list bullets, or formatting artifacts.

\paragraph{Voice selection.}
We synthesize the speech with the in-house \texttt{Chatterbox} variant described above.
For each side of a pair, we pick a reference voice in the correct language from the FLEURS speakers.
We pair a source-side voice with the most similar target-side voice, so that both languages are spoken by acoustically similar speakers.
We measure the similarity with \texttt{ECAPA-TDNN} speaker embeddings~\cite{ecapatdnn, speechbrain},\footnote{\href{https://huggingface.co/speechbrain/spkrec-ecapa-voxceleb}{\texttt{huggingface.co/speechbrain/spkrec-ecapa-voxceleb}}} and keep only voice pairs whose similarity is above a per-language threshold.

\paragraph{Speech synthesis and verification.}
We synthesize the source and target sentences with the selected voices.
Then, we transcribe every synthesized utterance with \texttt{Whisper} \texttt{Large-v3} \texttt{Turbo}\footref{fn-whisper} and align the transcript with the intended text.
This check detects trailing hallucinations and audio that continues after the intended text ends. 
We discard the utterances where the recognizer output differs from the reference; the remaining triples of source audio, target text, and target audio supervise MT, AST, and S2ST.

\subsubsection{Understanding supervision}
\label{subsec:understanding_supervision}
We construct two kinds of data for the understanding family.
Closed captioning data is synthesized by mixing labeled audio segments.
AudioReason-10M is a large corpus of reasoning-focused audio question answering.

\paragraph{Closed captioning: scene composition.}
Rich closed captions describe not only speech but also music, sound effects, speaker turns, and timing, yet no large corpus provides all of these annotations jointly.
We therefore build each audio scene from labeled segments and derive its caption from the known composition: every event, timestamp, and speaker label in the caption is exact with respect to the labels of the segments.

We draw segments from three libraries: speech, sound effects (SFX), and music.
Each segment has a label, a source (speaker) identifier, and a duration.
We sample segments with a simple transition model that alternates between speech and sound events; we place them one after another on a timeline, with randomized blank gaps and optional overlap.
Each scene lasts between 10 and 30 seconds, and we balance the corpus across four duration bands.

\paragraph{Closed captioning: speakers and acoustics.}
We control speaker composition and acoustic realism of each scene explicitly.
We can enforce a target number of distinct speakers per scene, drawing speech segments from speakers with enough utterances and encouraging natural speaker turns.
Segments are mixed with a separate intensity range for each type: speech is prominent, SFX are attenuated, and music is quiet.
We apply linear fade-in and fade-out and loudness normalization, and resize SFX segments at random to vary their duration.
Optionally, we add a quiet background music track over the whole scene.
The result is varied multi-speaker scenes with music and sound effects.

\paragraph{Closed captioning: caption formats.}
We generate several caption formats from the composition of each scene, one for each annotation task; a single scene therefore supervises several levels of detail.
The formats range from a transcription with bracketed non-speech events, to timestamped events, to timestamped and speaker-attributed captions.
The most detailed format is a structured summary, counting the music, speech, and SFX segments, listing the unique speakers, and describing each event with its label and time span, similar to~\cite{kumar2026tac}.
Prompts for each task are shown in Table~\ref{tab:prompts}.

\paragraph{Closed captioning: languages and stages.}
We cover Chinese, English, French, German, Italian, Japanese, and Spanish.
We use a separate segment library for each language, built from open-source~\cite{ardila2020common, defferrard2016fma, audioset, koudounas23_interspeech} and internal datasets.
We split this data across two stages: in Stage~2, we use only commercially licensed segment libraries; in Stage~3, we augment the mix with the rest.

\paragraph{AudioReason-10M: reasoning-focused audio QA.}
Audio QA built from recognition-oriented benchmarks, such as ASR or sound event classification, does not train expert-level auditory reasoning~\cite{ghosh2026audio}.
Following the approach of Audio Flamingo~3~\cite{ghosh2026audio}, we build a large reasoning-focused corpus, which we call AudioReason-10M.
It contains 10M audio QA pairs, most of them multiple-choice, on clips of at most 30 seconds.
We build it only from commercially licensed or internally owned audio, and we exclude research-only and license-restricted corpora.

We organize AudioReason-10M around five reasoning skill families, each requiring capabilities that go beyond simple transcription or single-label classification.
\begin{itemize}
    \item \textit{Temporal reasoning.} Questions require reasoning over the order, duration, and co-occurrence of events, for example which sound comes first, how long an event lasts, or whether two events overlap.
    \item \textit{Speech-in-sound.} The model must combine the spoken content with the surrounding acoustic scene, so that speech and non-speech evidence lead to one answer.
    \item \textit{Speaker reasoning.} Questions probe speaker attributes such as gender and approximate age, as well as the number of distinct speakers and their turns, requiring paralinguistic rather than lexical understanding.
    \item \textit{Spatial audio perception.} Questions ask about the direction, relative position, and movement of sound sources, which requires spatial cues in the audio.
    \item \textit{Instruction following with verifiable constraints.} Each question has an explicit output constraint, such as a format, a length, or an answer set, whose satisfaction can be checked automatically; this teaches the model to follow instructions precisely while reasoning about the audio.
\end{itemize}

For each skill, we design dedicated prompts and QA templates.
We generate distractors for the multiple-choice questions, with the aim of reducing shortcuts that do not need the audio.
All skills share one QA schema, so that we can mix them uniformly during training.

\begin{table}[!ht]
    \caption{\textbf{Performance comparison across all evaluated benchmarks.} For each dataset, we report the best prior system among comparable 4-10B models, including \texttt{Gemma4}, \texttt{Phi4-MM}, \texttt{AF3}, \texttt{Q2.5O} (Prior SOTA), the Commercial-Ready Walrus (\texttt{Walrus-CR}), and the Research Walrus (\texttt{Walrus-R}). Metrics: WER/CER ($\downarrow$) for ASR tasks, cpWER/cpCER ($\downarrow$) for MT-ASR, BLEU/ASR-BLEU ($\uparrow$) for AST/S2ST, accuracy ($\uparrow$) for IF and SER, CIDEr/CLAP-Sim ($\uparrow$) for captioning, R@1/5/10 ($\uparrow$) for retrieval. Averages over languages; per-language breakdowns are in Appendix~\secref{app:breakdowns}. \texttt{Phi4-MM} results on MLS and FLEURS are reported as averages across supported languages, i.e., with Ar/Hi/Ko/Nl/Pl excluded. Best results per dataset in \textbf{bold}.}
    \label{tab:master}
    \vspace{1mm}
    \centering
    \small
    \resizebox{\columnwidth}{!}{%
    \begin{tabular}{llcccc}
        \toprule
        \textbf{Task} & \textbf{Dataset} & \textbf{Metric} & \textbf{Prior SOTA} & \textbf{Walrus-CR} & \textbf{Walrus-R} \\
        \midrule
        \rowcolor{stdblue}
        \multicolumn{6}{@{}c}{\textit{Recognition}} \\
        ASR            & LS Clean/Other~\cite{librispeech}      & WER $\downarrow$     & \textbf{1.57}/\textbf{3.13} (AF3)   & 1.88/4.99    & 1.65/3.85 \\
        ASR            & MLS~\cite{pratap2020mls}                 & WER $\downarrow$     & 8.20 (Phi4-MM)                      & 9.14         & \wb{6.89} \\
        ASR            & FLEURS~\cite{conneau2023fleurs}              & WER $\downarrow$     & \textbf{4.00} (Phi4-MM)             & 7.54         & 5.33 \\
        ASR            & VoxPopuli~\cite{wang2021voxpopuli}           & WER $\downarrow$     & 11.78 (Gemma4)                      & 8.38         & \wb{7.84} \\
        ASR            & Internal Entertainment (avg)      & WER $\downarrow$     & 24.35 (Gemma4)                        & 26.71        & \wb{17.45} \\
        CS-ASR         & CS-FLEURS Read/XTTS~\cite{yan25c_interspeech} & CER $\downarrow$     & 32.30/44.10 (Gemma4)                & 26.69/35.36  & \wb{7.23}/\wb{15.01} \\
        MT-ASR         & LibriMix 2Mix~\cite{cosentino2020librimix}       & cpWER $\downarrow$   & N/A                                 & 26.84        & \wb{19.32} \\
        MT-ASR         & JA Daily Talk~\cite{cejc}       & cpCER $\downarrow$   & N/A                                 & 66.60        & \wb{56.29} \\
        \midrule
        \rowcolor{specgreen}
        \multicolumn{6}{@{}c}{\textit{Translation}} \\
        AST X$\to$En   & CoVoST-2~\cite{wang2020covost}/FLEURS~\cite{conneau2023fleurs}    & BLEU $\uparrow$       & 39.33/29.86 (Phi4-MM)               & 40.28/30.21   & \wb{42.63}/\wb{33.05} \\
        AST En$\to$X   & FLEURS~\cite{conneau2023fleurs}             & BLEU $\uparrow$       & 32.15 (Phi4-MM)                     & 32.42         & \wb{34.33} \\
        S2ST X$\to$En  & CVSS-C~\cite{jia2022cvss}/FLEURS~\cite{conneau2023fleurs}      & ASR-BLEU $\uparrow$   & 36.02/25.15 (Q25O)                  & 43.27/28.59   & \wb{43.63}/\wb{32.28} \\
        S2ST En$\to$X  & FLEURS~\cite{conneau2023fleurs}             & ASR-BLEU $\uparrow$   & N/A                                 & 17.19         & \wb{22.94} \\
        \midrule
        \rowcolor{accentyellow}
        \multicolumn{6}{@{}c}{\textit{Understanding}} \\
        IF             & MMAU-Pro~\cite{kumar2026mmau}           & ACC $\uparrow$        & 52.20 (Q25O)                        & 50.18         & \wb{71.92} \\
        IF             & MMSU~\cite{wang2026mmsu}               & ACC $\uparrow$        & 61.40 (AF3)                         & 55.18         & \wb{64.14} \\
        SER            & IEMOCAP~\cite{busso2008iemocap}            & ACC $\uparrow$        & \textbf{63.80} (AF3)                & 41.44         & 49.55 \\
        Captioning     & Clotho-v2~\cite{clotho}          & CIDEr/CLAP-Sim $\uparrow$ & \textbf{0.50}/\textbf{0.41} (AF3) & 0.10/0.19   & 0.39/0.35 \\
        Retrieval A2T  & AudioCaps~\cite{audiocaps}          & R@1/5/10 $\uparrow$   & \textbf{0.20}/0.60/0.76 (AF3)       & 0.16/0.52/0.67 & 0.18/\wb{0.63}/\wb{0.78} \\
        Retrieval T2A  & AudioCaps~\cite{audiocaps}          & R@1/5/10 $\uparrow$   & 0.21/0.59/0.76 (AF3)                & 0.16/0.48/0.65 & \wb{0.22}/\wb{0.60}/\wb{0.78} \\
        \bottomrule
    \end{tabular}
    }
\end{table}

\section{Results}
\label{sec:results}
We evaluate Walrus on recognition, translation, and understanding benchmarks.
Table~\ref{tab:master} compares Walrus with the best prior system among comparable 4-10B models: Gemma 4 E4B (\texttt{Gemma4})\footref{fn-gemma4-e4b}~\cite{team2026gemma}, Phi-4-Multimodal (\texttt{Phi4-MM})\footnote{\href{https://huggingface.co/microsoft/Phi-4-multimodal-instruct}{\texttt{huggingface.co/microsoft/Phi-4-multimodal-instruct}}}~\cite{phi4-mm}, Audio-Flamingo-3 (\texttt{AF3})\footnote{\href{https://huggingface.co/nvidia/audio-flamingo-3}{\texttt{huggingface.co/nvidia/audio-flamingo-3}}}~\cite{ghosh2026audio}, and Qwen2.5-Omni (\texttt{Q2.5O})\footnote{\href{https://huggingface.co/Qwen/Qwen2.5-Omni-7B}{\texttt{huggingface.co/Qwen/Qwen2.5-Omni-7B}}}~\cite{xu2025qwen2}.
We report both variants, the Commercial-Ready model (\texttt{Walrus-CR}, Stage 2) and the Research model (\texttt{Walrus-R}, Stage 3).
All scores are averaged over languages; per-language breakdowns are in Appendix~\secref{app:breakdowns}, where we also compare our models with larger (Qwen3-Omni-30B (\texttt{Q3O}) \cite{xu2025qwen3}) or dedicated systems, such as Whisper Large-v3 (\texttt{WLv3}), SeamlessM4T v2 (\texttt{Seamless}) \cite{barrault2023seamless}, and VibeVoice-ASR~\cite{peng2026vibevoice}.

\paragraph{Multilingual and English ASR.}
On English LibriSpeech, \texttt{Walrus-R} reaches 1.65/3.85 WER on clean/other, slightly behind \texttt{AF3} (1.57/3.13).
On multilingual benchmarks, it has the lowest WER on MLS (6.89 against 8.20 for \texttt{Phi4-MM}, Table~\ref{tab:mls}) and on VoxPopuli (7.84 against 11.78 for \texttt{Gemma4}, Table~\ref{tab:voxpopuli}).
On FLEURS (Table~\ref{tab:fleurs-asr}), it reaches 5.33 WER (94.7\% word accuracy, Figure~\ref{fig:radar}); this is ahead of the 4$\times$-bigger Qwen3-Omni (93.8\%) and Whisper Large-v3 (92.0\%).
\texttt{Phi4-MM} reaches 4.00 WER, but it does not support five of our evaluation languages (Arabic, Dutch, Hindi, Korean, and Polish; see Table~\ref{tab:fleurs-asr}).
On internal entertainment audio (Table~\ref{tab:asr-internal}), \texttt{Walrus-R} reaches 17.45 WER, against 24.35 for \texttt{Gemma4}.

\paragraph{Code-switching ASR.}
On CS-FLEURS read speech, \texttt{Walrus-R} reaches 7.23 CER, against 32.30 for the best prior system (\texttt{Gemma4}).
On the XTTS subset, it reaches 15.01 CER, against 44.10.
\texttt{Walrus-CR} reaches 26.69 and 35.36, which is already better than every prior system.
Both systems outperform even the 30B \texttt{Q3O} model, which achieves only 38.70 CER on read speech (Table~\ref{tab:cs-read}).

\paragraph{Multi-talker ASR.}
Table~\ref{tab:mt-librimix-jdailytalk} shows the results on overlapping speech.
\texttt{Gemma4}, \texttt{Phi4-MM}, \texttt{AF3}, and \texttt{Q2.5O} cannot transcribe overlapping speakers, and they produce no valid output.
On LibriMix clean 2-mix, \texttt{Walrus-R} reaches 19.32 cpWER; this is about half of the 37.04 of Qwen3-Omni, and far below the 59.34 of the dedicated VibeVoice-ASR.
On Japanese Daily Talk, it reaches 56.29 cpCER, against 71.99 for VibeVoice-ASR.
\texttt{Walrus-CR} also beats VibeVoice-ASR on the \textit{cp} metrics.
The time-constrained metrics (tcpWER and tcpCER) give the same ranking for \texttt{Walrus-R}.
Two points limit this comparison: (i) we evaluate on 30-second segments, which match the input limit of Walrus but not the long-form design of VibeVoice-ASR; and (ii) LibriMix is also part of our training data, so it is in-domain for Walrus.
The absolute error on Japanese Daily Talk is still high, thus showing how natural multi-speaker conversation still remains a hard task.

\paragraph{Speech translation.}
On AST, \texttt{Walrus-R} reaches 42.63 BLEU on CoVoST-2 (Table~\ref{tab:ast-covost}) and 33.05 on FLEURS X$\to$En (Table~\ref{tab:ast-fleurs-xen-enx}), against 39.33 and 29.86 for \texttt{Phi4-MM}.
On FLEURS En$\to$X, it reaches 34.33 BLEU, against 32.15.
On S2ST, Walrus is the strongest system: it achieves 43.63 ASR-BLEU on CVSS-C X$\to$En and 32.28 on FLEURS X$\to$En, against 36.02 and 25.15 for Qwen2.5-Omni.
On CVSS-C (Table~\ref{tab:s2st-cvss}), it also surpasses the 30B Qwen3-Omni, while being close on FLEURS (Table~\ref{tab:s2st-fleurs-xen}).
It also supports En$\to$X S2ST (22.94 ASR-BLEU on FLEURS), where the other systems report no result.
\texttt{Walrus-CR} is already ahead of the prior systems on AST and S2ST X$\to$En, but its gains are smaller on both tasks.

\paragraph{Audio understanding.}
\texttt{Walrus-R} is the strongest system on audio understanding.
On MMAU-Pro, it reaches 71.92\% accuracy, far above the best baseline (52.20\% for Qwen2.5-Omni), and also closed-source systems such as for Gemini-Flash-3.7 (70.40\%), and GPT-4o-Audio (52.50\%).
On MMSU, it reaches 64.14\%, against 61.40\% for \texttt{AF3}.
On captioning and retrieval, Walrus is competitive with the best specialized baseline (\texttt{AF3}), leading on several AudioCaps retrieval metrics and closing most of the gap on Clotho-v2 captioning.
It trails \texttt{AF3} on IEMOCAP emotion recognition (49.55\% against 63.80\%, OOD for our model), which reflects the small share of emotion data in our mix.

\paragraph{Commercial-Ready and Research variants.}
\texttt{Walrus-R} improves on \texttt{Walrus-CR} on every benchmark in Table~\ref{tab:master}.
The largest gains are on CS-FLEURS read speech (26.69 to 7.23 CER), internal entertainment ASR (26.71 to 17.45 WER), and MMAU-Pro (50.18\% to 71.92\%).
The two variants share the backbone, but they differ in the training stage, data license, and task weights (Sec.~\secref{sec:variants}).
The gains therefore reflect Stage 3 as a whole, and we do not separate the effect of each change.

\section{Discussion}
\label{sec:discussion}
Walrus uses a standard backbone, so we discuss its results in terms of the data.
We highlight three findings that the benchmark tables alone do not make explicit.

\paragraph{Targeted data goes with targeted gains.}
The largest gains appear where we built dedicated data.
CS-ASR is only 1.1\% of the recognition data, which is about 0.5\% of all data.
Still, the CER on CS-FLEURS read speech falls from 32.30 for the best prior system to 26.69 for \texttt{Walrus-CR} and 7.23 for \texttt{Walrus-R}.
Entertainment-domain ASR follows the same pattern: the internal entertainment WER falls from 26.71 to 17.45 between the two variants, and the Japanese subset falls from 23.69 to 12.90 (Appendix~\secref{app:breakdowns}, Table~\ref{tab:asr-internal}).
MMAU-Pro accuracy rises from 50.18\% to 71.92\% between the two variants.

\paragraph{Weighting and recognition.}
Stage 3 raises the weights of generation and understanding and lowers the ASR weight from 5.0 to 4.0.
Even so, \texttt{Walrus-R} improves on \texttt{Walrus-CR} on every ASR benchmark.
This is consistent with the idea that a high recognition weight protects the core while the mix shifts.
However, as we did not run Stage 3 data with the Stage 2 weights, we cannot fully separate the effect of the weights from the effect of the new data.

\paragraph{Remaining gaps.}
Walrus trails the best baseline on emotion recognition and audio captioning.
On the former, it reaches 49.55\% on IEMOCAP (OOD) against 63.80\% for \texttt{AF3}. This partly reflect the distribution of SER data in our mix (about 0.3\%).
On audio captioning, it achieves 0.39 CIDEr on Clotho-v2 (OOD) against 0.50 (\texttt{AF3}), while captioning represents 15.4\% of the understanding data.
Furthermore. on Japanese Daily Talk, \texttt{Walrus-R} reaches 56.29 cpCER, which beats VibeVoice-ASR but is still high in absolute terms.
Our multi-talker training data is mostly simulated, and real overlapping speech is harder.

\section{Conclusion}
\label{sec:conclusion}
We presented Walrus, a unified 8B speech-and-audio model that covers recognition, translation, generation, and understanding.
Its core is a three-stage curriculum, from broad multilingual data to complex and domain-specific data. Task-aware and dataset-aware weighting balances the very different token budgets of the tasks.
Trained on over 600M samples across more than 20 tasks and 100 languages, \textit{Walrus-R} matches or surpasses comparable open SpeechLLMs on most benchmarks, and leads clearly on code-switching ASR, multilingual and entertainment-domain recognition, speech-to-speech translation, and audio understanding.
On several benchmarks, it also outperforms larger or dedicated systems such as Qwen3-Omni-30B, Seamless v2, and VibeVoice-ASR.
These results suggest that data engineering is a primary driver of quality in multi-task speech and audio modeling.

\bibliographystyle{abbrv}
\bibliography{mybib}


\newpage

\appendix

\section{Task prompts}
\label{app:prompts}
Walrus is steered to each of its tasks only through a natural-language prompt.
This appendix describes how we design, sample, and use these prompts.
Table~\ref{tab:prompts} lists the canonical prompt of each task.

\subsection{Prompt pools}
\label{app:prompt-pools}
We do not use a fixed instruction for each task, because the model could overfit to a single phrasing.
Instead, we keep a pool of 10-20 paraphrased prompts for every task, and we sample one prompt from the pool for each training example.
This encourages robust instruction following, since the model must read the instruction instead of recognizing a fixed string.

The prompts in a pool differ in register and in format.
\begin{itemize}
    \item \textit{Imperative prompts} give a direct command, for example \textit{``Transcribe the {\{lang1\}} speech:''}.
    \item \textit{Interrogative prompts} ask a question, for example \textit{``What is being said in the speech?''}.
    \item \textit{Tag-style prompts} use a compact marker, for example \textit{``[ASR {\{lang1\}}]''}.
    \item \textit{Structured prompts} use explicit delimiters, for example \texttt{<context>...</context>}.
\end{itemize}
All prompts of a task ask for the same output; only the wording and the format change, so the paraphrases never change the task itself.

\subsection{Slot variables}
\label{app:prompt-slots}
Prompts are templates with slot variables (Table~\ref{tab:prompt-slots}); we fill such slots with the values of each sample when building the example. One template thus serves many languages, contexts, and label sets.

\begin{table}[!ht]
    \centering
    \caption{\textbf{Slot variables used in the prompt templates.}}
    \label{tab:prompt-slots}
    \vspace{1mm}
    \small
    \begin{tabular}{ll}
        \toprule
        \textbf{Slot} & \textbf{Filled with} \\
        \midrule
        \texttt{\{lang1\}}, \texttt{\{lang2\}} & Source and target language \\
        \texttt{\{context\}} & Preceding text, as discourse context \\
        \texttt{\{keywords\}} & List of candidate keywords \\
        \texttt{\{labels\}} & Closed set of allowed labels, for example emotions \\
        \texttt{\{speaker\}} & Target speaker for target-speaker queries \\
        \texttt{\{start\}}, \texttt{\{end\}} & Time span for time-conditioned queries \\
        \texttt{\{agent\}}, \texttt{\{user\}} & Roles in dialogue-level context \\
        \bottomrule
    \end{tabular}
    \vspace{-3mm}
\end{table}

\subsection{Sampling and canonical prompts}
\label{app:prompt-canonical}
We sample uniformly from each pool, except for one prompt that we mark as canonical, for which we give a higher sampling probability. 
Table~\ref{tab:prompts} lists these canonical prompts of each task.

Each pool has a \texttt{fixed} flag, which selects the canonical prompt deterministically.
We use this flag at evaluation time, so that results do not depend on random prompt choice and can be reproduced.

\begin{table}[!ht]
    \centering
    \caption{\textbf{Overview of the data pipelines.}
    ``Tools'' lists the models and recipes used, ``Quality control'' the filters applied to the output.
    A dash means that we apply no filter beyond the construction itself.}
    \label{tab:data-pipelines}
    \vspace{1mm}
    \footnotesize
    \setlength{\tabcolsep}{3pt}
    \renewcommand{\arraystretch}{1.15}
    \begin{tabular}{
        >{\raggedright\arraybackslash}p{0.14\linewidth}
        >{\raggedright\arraybackslash}p{0.22\linewidth}
        >{\raggedright\arraybackslash}p{0.27\linewidth}
        >{\raggedright\arraybackslash}p{0.27\linewidth}}
        \toprule
        \textbf{Pipeline} & \textbf{What we build} & \textbf{Tools} & \textbf{Quality control} \\
        \midrule
        \rowcolor{stdblue}
        \multicolumn{4}{@{}c}{\textit{Transcript cleaning}} \\
        Casing and punctuation & Restored transcripts of real ASR corpora & Qwen3.5-9B with few-shot prompts & CER $\leq$ 0.05 against the original; length check \\
        Text normalization & Consistent references across languages & Rules for European, CJK, and Hindi/Arabic scripts & Manual inspection of samples \\
        \midrule
        \rowcolor{stdblue}
        \multicolumn{4}{@{}c}{\textit{Recognition}} \\
        Context ASR & Target utterance with 1-5 preceding utterances & MLS chapters, internal domains & -- \\
        Keyword ASR & Keyword lists of 0-100 items with distractors & Named-entity recognition & Entities in at least two utterances \\
        Code-switching ASR & Synthetic text and audio, 13 language pairs & Gemma 4 26B, Qwen3-TTS, Chatterbox (internal), FLEURS reference voices & De-duplication; NISQA threshold; Whisper WER/CER threshold \\
        Multi-talker ASR & Real meeting clips and simulated 2- to 4-speaker mixtures & DiCoW recipe, FastMSS, Qwen3-ForcedAligner & Clips of at most 30 seconds \\
        Entertainment speech & Curated web speech and augmented corpora & YODAS-Granary, audio effects & Same cleaning pipeline as other ASR data \\
        \midrule
        \rowcolor{specgreen}
        \multicolumn{4}{@{}c}{\textit{Translation and dialogues}} \\
        Synthetic dialogues & Multi-turn dialogues with context and keyword tasks & Gemma 4 26B, Qwen3.5-9B, Chatterbox variant, Whisper, COMET-Kiwi or LLM judge & Judge ensemble; error rate $\leq$ 0.1; quality score $\geq$ 0.80 \\
        Parallel text and speech & MT, AST, and S2ST triples from OPUS & multilingual e5, Chatterbox (internal), ECAPA-TDNN, Whisper & Top decile by similarity; voice-similarity threshold; Whisper check \\
        \midrule
        \rowcolor{accentyellow}
        \multicolumn{4}{@{}c}{\textit{Understanding}} \\
        Closed captioning & Mixed scenes of 10-30 seconds with exact captions & Speech, SFX, and music segment libraries & Several constraints enforced during creation, manual inspection of early samples \\
        AudioReason-10M & 10M reasoning QA pairs, five skills & Dedicated prompts and QA templates & Commercially licensed or internal audio only; constraints enforced in QA synthesis \\
        \bottomrule
    \end{tabular}
    \vspace{-3mm}
\end{table}

\subsection{Prompts for conditioned tasks}
\label{app:prompt-conditioned}
Many tasks need more than a language name, so their prompts carry extra inputs.
The structure of these inputs is the same across the variants of a task.
\begin{itemize}
    \item \textit{Context ASR and context AST/S2ST.} The prompt contains the preceding text in a marked block, such as \texttt{<context>...</context>}.
    For dialogues, the source and target sides of earlier turns are marked, and the roles come from \texttt{\{agent\}} and \texttt{\{user\}}.
    \item \textit{Keyword tasks.} The prompt contains a shuffled list of candidate keywords.
    The list may contain distractors and may be empty, as described in Sec.~\secref{sec:recognition-supervision}.
    \item \textit{Timestamp and multi-talker tasks.} The prompt states the requested output, such as speaker labels or timestamps.
    For target-speaker ASR and timestamp QA, it also contains the target speaker or the time span.
    \item \textit{Emotion recognition.} The prompt lists the allowed labels with \texttt{\{labels\}}, so that the answer is one of a closed set.
    \item \textit{Closed captioning.} Each level of detail has its own prompt, from a plain transcription with non-speech events to the full structured summary (Sec.~\secref{subsec:understanding_supervision}).
    \item \textit{Translation and generation.} The prompt names the source and target language.
    It also states whether the output is text or speech, and whether the translation is direct or cascaded.
\end{itemize}

\subsection{Prompts at inference}
\label{app:prompt-inference}
At inference time, a user can write a prompt in any form, not only the canonical one.
We report all results with the canonical prompts of Table~\ref{tab:prompts}.
Since each task is trained with many paraphrases, we expect the model to follow new wordings; however, we do not measure this robustness in this report.

\begin{table}[!ht]
  \caption{\textbf{ASR Results: LibriSpeech (WER $\downarrow$).} \texttt{WLv3} indicates Whisper-Large-v3, \texttt{AF3} Audio-Flamingo3, \texttt{Q250/Q30} denote Qwen2.5Omni and Qwen3Omni, respectively. Best results in \textbf{bold}, second-best \underline{underlined.}}
  \label{tab:ls}
  \vspace{1mm}
  \centering
  \begin{tabular}{lcc}
    \toprule
    \textbf{Model} & \textbf{Clean} & \textbf{Other} \\
    \midrule
    \rowcolor{nonllmgray}
    WLv3        & 2.31 & 4.29 \\
    Gemma4      & 3.22 & 7.74 \\
    Phi4-MM     & 1.68 & 3.83 \\
    AF3         & \textbf{1.57} & \textbf{3.13} \\
    Q25O        & 1.83 & \underline{3.44} \\
    \rowcolor{q3ored}
    Q3O         & 1.22 & 2.48 \\
    \midrule
    Walrus-CR        & 1.88 & 4.99 \\
    Walrus-R         & \underline{1.65} & 3.85 \\
    \bottomrule
  \end{tabular}
\end{table}

\begin{table}[!ht]
  \caption{\textbf{ASR Results: MLS (WER $\downarrow$).} *This average score reflects only the supported languages.}
  \label{tab:mls}
  \vspace{1mm}
  \centering
  \resizebox{0.9\columnwidth}{!}{%
  \begin{tabular}{lcccccccccc}
    \toprule
    \textbf{Model} & \textbf{DE} & \textbf{EN} & \textbf{ES} & \textbf{FR} & \textbf{IT} & \textbf{NL} & \textbf{PL} & \textbf{PT} & \textbf{AVG} \\
    \midrule
    \rowcolor{nonllmgray}
    WLv3        & 5.54  & 5.30 & 4.01  & 4.60  & 9.09  & 10.36 & 4.89  & 7.20 & 6.37 \\
    Gemma4      & 8.71  & 9.15 & 5.43 & 7.86  & 14.63 & 12.06 & \underline{10.44} & 9.10 & 9.67 \\
    Phi4-MM     & 7.59  & 8.04 & 10.49 & \underline{5.15} & \underline{11.63} & N/A   & N/A   & \textbf{6.29} & \underline{8.20}$^{*}$ \\
    AF3         & 27.45 & \textbf{5.02} & 12.57 & 20.07 & 19.96 & 28.46 & 53.50 & 17.26 & 23.04 \\
    Q25O        & \underline{6.54}  & 6.14 & 8.21  & 7.92  & 20.27 & 17.56 & 19.21 & 11.96 & 12.23 \\
    \rowcolor{q3ored}
    Q3O         & 3.77  & 3.77 & 3.41  & 3.00  & 9.72  & 8.94  & 12.06 & 5.22 & 6.24 \\
    \midrule
    Walrus-CR        & 7.88  & 6.43 & \underline{4.44} & 5.67  & 14.52 & \underline{11.53} & 13.63 & 9.02 & 9.14 \\
    Walrus-R         & \wb{5.88} & \underline{5.91} & \wb{3.47}  & \wb{4.46}  & \wb{9.86}  & \wb{10.22} & \wb{7.71}  & \underline{7.67} & \wb{6.89} \\
    \bottomrule
  \end{tabular}
  }
\end{table}

\section{Summary of data pipelines}
\label{app:data-summary}
Sec.~\secref{sec:data-prep} describes the data pipelines of Walrus in five subsections; this appendix gives a short overview of all of them in one place.
Table~\ref{tab:data-pipelines} lists, for each pipeline, what we build, which tools we use, and which quality controls we apply.
The rows follow the order of Sec.~\secref{sec:data-prep}.

\paragraph{How to read the table.}
The column \textit{``What we build''} names the supervision that the pipeline produces;
the column \textit{``Tools''} names the corpora, models, and recipes that the pipeline uses;
and the column \textit{``Quality control''} names the filters that we apply to the output.
A dash means that we apply no filter beyond the construction itself.

\paragraph{Real and synthetic data.}
Our pipelines use real and synthetic data in different ways.
Some pipelines only restructure real data: context ASR and keyword ASR add context or keyword inputs to real ASR corpora, and the casing restoration changes only the text surface.
Some pipelines recombine real audio: simulated multi-talker mixtures rearrange real single-speaker utterances, and closed-captioning scenes mix real speech, sound-effect, and music segments.
Other pipelines generate new content: code-switching ASR and the synthetic dialogues generate both text and audio, and the translation pipeline synthesizes speech for mined text pairs, while AudioReason-10M builds new questions and answers over audio.

This difference matters for the reliability of the labels.
In recombined data, the labels follow from the construction, but they are only as accurate as the labels of the source segments.
In generated data, the text is the label, but the audio can differ from it if the TTS system makes an error. We therefore check the audio of generated data, as described next.

\begin{table}[!ht]
  \caption{\textbf{ASR Results: FLEURS (WER/CER $\downarrow$).} *This average score reflects only the supported languages.}
  \label{tab:fleurs-asr}
  \vspace{1mm}
  \centering
  \resizebox{\columnwidth}{!}{%
  \begin{tabular}{lcccccccccccc}
    \toprule
    \textbf{Model} & \textbf{AR} & \textbf{DE} & \textbf{EN} & \textbf{ES} & \textbf{FR} & \textbf{HI} & \textbf{IT} & \textbf{JA} & \textbf{KO} & \textbf{PT} & \textbf{ZH} & \textbf{AVG} \\
    \midrule
    \rowcolor{nonllmgray}
    WLv3        & 16.40 & 5.96 & 6.24 & 4.26 & 6.96 & 18.61 & 4.07 & 6.55 & 4.34 & 5.55 & 9.03 & 7.99 \\
    Gemma4      & 16.20 & 6.10 & 6.50 & 3.50 & 8.00 & \underline{8.60} & 3.20 & 7.80 & \underline{5.30} & 4.60 & 13.60 & 7.58 \\
    Phi4-MM     & N/A & \textbf{3.96} & \textbf{3.38} & \underline{3.02} & \textbf{4.35} & N/A & \textbf{1.98} & \textbf{4.50} & N/A & \underline{3.98} & \textbf{6.83} & \textbf{4.00}$^{*}$ \\
    AF3         & N/A & 22.53 & 5.06 & 25.83 & 39.82 & 74.45 & 19.74 & N/A & 54.65 & 49.65 & N/A & 36.47 \\
    Q25O        & 18.75 & 7.67 & 6.46 & 5.70 & 5.08 & 14.54 & 7.63 & 8.12 & 7.03 & 6.50 & \underline{7.29} & 8.62 \\
    \rowcolor{q3ored}
    Q3O         & 14.84 & 5.02 & 7.36 & 4.27 & 5.10 & 7.59  & 3.94 & 4.85 & 3.31 & 5.25 & 6.78  & 6.21 \\
    \midrule
    Walrus-CR        & \underline{13.48} & 6.03 & 5.49 & 4.16 & 8.51 & 10.51 & 4.06 & 7.99 & 7.37 & 5.03 & 10.35 & 7.54 \\
    Walrus-R         & \wb{10.56} & \underline{4.34} & \underline{4.50} & \wb{2.94} & \underline{4.84} & \wb{4.94} & \underline{2.55} & \underline{7.64} & \wb{4.84} & \wb{3.87} & 7.70 & \underline{5.33} \\
    \bottomrule
  \end{tabular}
  }
\end{table}

\begin{table}[!ht]
  \caption{\textbf{ASR Results: VoxPopuli (WER $\downarrow$).} 5 major EU languages are considered.}
  \label{tab:voxpopuli}
  \vspace{1mm}
  \centering
  \begin{tabular}{lcccccc}
    \toprule
    \textbf{Model} & \textbf{DE} & \textbf{EN} & \textbf{ES} & \textbf{FR} & \textbf{IT} & \textbf{AVG} \\
    \midrule
    \rowcolor{nonllmgray}
    WLv3        & 18.26  & 9.89 & 10.19 & 9.98  & 27.21  & 15.11 \\
    Gemma4      & 13.76  & 8.67 & 8.79  & 10.89 & 16.80  & 11.78 \\
    Phi4-MM     & 12.13  & 5.91 & 10.18 & 13.18 & 22.67  & 12.81 \\
    AF3         & 108.73 & 5.55 & 95.36 & 91.02 & 142.65 & 86.66 \\
    Q25O        & 15.84  & 6.64 & 13.24 & 11.39 & 23.67  & 14.16 \\
    \rowcolor{q3ored}
    Q3O         & 11.60  & 7.66 & 8.35  & 8.06  & 15.73  & 10.28 \\
    \midrule
    Walrus-CR        & \underline{7.54} & \wb{4.37} & \underline{5.72} & \underline{9.15} & \underline{15.14} & \underline{8.38} \\
    Walrus-R         & \wb{6.77} & \underline{4.82} & \wb{4.95} & \wb{8.42} & \wb{14.23} & \wb{7.84} \\
    \bottomrule
  \end{tabular}
\end{table}

\begin{table}[!ht]
  \caption{\textbf{ASR Results: CoVoST-2 (WER/CER $\downarrow$).} *This average score reflects only the supported languages.}
  \label{tab:covost-asr}
  \vspace{1mm}
  \centering
  \resizebox{\columnwidth}{!}{%
  \begin{tabular}{lccccccccc}
    \toprule
    \textbf{Model} & \textbf{DE} & \textbf{ES} & \textbf{FR} & \textbf{IT} & \textbf{JA} & \textbf{PT} & \textbf{RU} & \textbf{ZH} & \textbf{AVG} \\
    \midrule
    \rowcolor{nonllmgray}
    WLv3        & 6.78  & 4.19  & 12.28 & 5.10  & 10.46  & 4.52  & 4.21  & 16.37  & 7.99 \\
    Gemma4      & 9.33  & 5.55  & 14.58 & 7.10  & 16.46  & 8.16  & 9.16  & 47.37  & 14.71 \\
    Phi4-MM     & 10.82 & \textbf{3.57}  & \textbf{6.05}  & \textbf{3.53}  & 36.91  & 10.26 & N/A   & 48.29  & 17.49$^{*}$ \\
    AF3         & 28.22 & 14.52 & 22.27 & 18.23 & 139.96 & 18.72 & 95.60 & 143.12 & 60.08 \\
    Q25O        & \underline{6.67}  & 7.51  & 8.44  & 8.62  & 19.85  & 12.88 & 7.95  & \textbf{10.69}  & 10.33 \\
    \rowcolor{q3ored}
    Q3O         & 4.71  & 4.26  & 6.91  & 4.37  & 8.87   & 6.83  & 3.51  & 4.70   & 5.52 \\
    \midrule
    Walrus-CR        & 7.01  & 4.12  & 8.43  & 5.21  & \underline{11.61}  & \underline{5.70}  & \underline{5.28}  & 18.49  & \underline{8.23} \\
    Walrus-R         & \wb{5.81} & \underline{3.62} & \underline{7.35}  & \underline{4.09}  & \wb{10.30}  & \wb{4.74}  & \wb{3.78}  & \underline{11.34}  & \wb{6.38} \\
    \bottomrule
  \end{tabular}
  }
\end{table}

\paragraph{Quality control.}
The quality controls in Table~\ref{tab:data-pipelines} fall into three groups.
\begin{itemize}
    \item \textit{Text filters.}
    These filters check the text before any audio is made.
    Examples are the CER check after casing restoration (Sec.~\secref{sec:transcript-cleaning}), the de-duplication of code-switched sentences, the LLM-judge ensemble for dialogues (Sec.~\secref{sec:conversational-supervision}), and the top-decile similarity filter for parallel text (Sec.~\secref{sec:translation-supervision}).
    \item \textit{Audio filters.}
    These filters check the synthesized audio itself.
    Examples are the NISQA threshold for code-switching speech and the speaker-similarity threshold for voice pairs.
    \item \textit{Text-audio checks.}
    These filters check that the audio says the intended text.
    We transcribe the synthesized speech with \texttt{Whisper} \texttt{Lv3-Turbo} and compare the transcript with the intended text; we use this check for code-switching speech, dialogues, and translation speech.
\end{itemize}
Translation quality of the dialogues is checked in a fourth way.
We score the target text with a reference-free estimator: an LLM judge in Stage~2 and COMET-Kiwi in Stage~3 (Sec.~\secref{sec:conversational-supervision}).

\paragraph{Models used to build the data.}
The tools in Table~\ref{tab:data-pipelines} play four roles.
\begin{itemize}
    \item \textit{Text generation and judging.} \texttt{Gemma} \texttt{4} \texttt{26B} and \texttt{Qwen3.5-9B} write code-switched sentences and dialogues, restore casing, and act as judges; \texttt{Qwen3.5-27B} translates dialogues from English onto different languages.
    \item \textit{Speech synthesis.} \texttt{Qwen3-TTS} and a fast in-house variant of \texttt{Chatterbox} create the audio for code-switching, dialogues, and translation.
    \item \textit{Recognition and alignment.} \texttt{Whisper} \texttt{Lv3-Turbo} checks synthesized speech, and \texttt{Qwen3-ForcedAligner} provides word-level alignments for multi-talker mixtures.
    \item \textit{Scoring.} \texttt{NISQA} scores speech quality, \texttt{ECAPA-TDNN} scores speaker similarity, \texttt{multi-} \texttt{lingual-e5} scores cross-lingual text similarity, and \texttt{COMET-Kiwi} scores translation quality.
\end{itemize}

\begin{table}[!ht]
  \caption{A\textbf{SR Results: Internal entertainment data (WER/CER $\downarrow$)}. Per-language averages over different test sets.}
  \label{tab:asr-internal}
  \vspace{1mm}
  \centering
  \begin{tabular}{lccccccc}
    \toprule
    \textbf{Model} & \textbf{DE} & \textbf{EN-UK} & \textbf{EN-US} & \textbf{ES} & \textbf{FR} & \textbf{IT} & \textbf{JA} \\
    \midrule
    \rowcolor{nonllmgray}
    WLv3        & 22.33 & 8.31  & 11.10 & 11.67 & 22.53 & 13.56 & 24.77 \\
    Gemma4      & 39.20 & 16.86 & 17.99 & \underline{17.72} & \underline{34.96} & \underline{18.35} & 25.37 \\
    Phi4-MM     & 69.58 & 17.44 & 20.25 & 26.91 & 66.01 & 22.40 & 202.35 \\
    AF3         & 63.90 & 16.05 & 17.74 & 47.20 & 88.49 & 41.51 & 169.11 \\
    Q25O        & \underline{34.54} & 18.55 & 22.72 & 30.55 & 37.68 & 28.55 & 43.18 \\
    \rowcolor{q3ored}
    Q3O         & 27.11 & 11.85 & 11.14 & 15.50 & 21.95 & 13.00 & 25.88 \\
    \midrule
    Walrus-CR   & 39.71 & \underline{15.94} & \underline{17.42} & 30.62 & 59.16 & 24.20 & \underline{23.69} \\
    Walrus-R    & \wb{27.68} & \wb{14.53} & \wb{14.26} & \wb{15.15} & \wb{24.55} & \wb{14.01} & \wb{11.97} \\
    \bottomrule
  \end{tabular}
  \vspace{-3mm}
\end{table}

\begin{table}[!ht]
  \caption{\textbf{CS-ASR Results: CS-FLEURS Read (top) and XTTS (bottom) test sets, CER ($\downarrow$).} *This average score reflects only the supported languages.}
  \label{tab:cs-read}
  \vspace{1mm}
  \centering
  \begin{tabular}{lcccccc}
    \toprule
    \textbf{Model} & \textbf{EN-HI} & \textbf{EN-JA} & \textbf{EN-DE} & \textbf{EN-FR} & \textbf{EN-ZH} & \textbf{AVG} \\
    \midrule
    \multicolumn{7}{@{}c}{\textit{Read Test Set}} \\
    \midrule
    \rowcolor{nonllmgray}
    WLv3        & 42.75 & 35.41 & 6.10  & 12.21 & 17.58 & 22.81 \\
    Gemma4      & 47.50 & 48.07 & 19.18 & 18.17 & 28.56 & 32.30 \\
    Phi4-MM     & N/A   & 69.29 & 29.15 & 32.78 & 81.71 & 53.23$^{*}$ \\
    AF3         & 97.95 & 85.09 & 64.50 & 73.62 & 86.69 & 81.57 \\
    Q25O        & 56.96 & 63.68 & 41.66 & 37.21 & 58.42 & 51.58 \\
    \rowcolor{q3ored}
    Q3O         & 35.74 & 66.47 & 38.64 & 19.41 & 33.02 & 38.66 \\
    \midrule
    Walrus-CR   & \underline{28.80} & \underline{44.84} & \underline{15.52} & \underline{17.80} & \underline{26.48} & \underline{26.69} \\
    Walrus-R    & \wb{7.88} & \wb{12.91} & \wb{3.63} & \wb{3.68} & \wb{8.05} & \wb{7.23} \\
    \midrule \midrule
    \multicolumn{7}{@{}c}{\textit{XTTS Test Set}} \\
    \midrule
    \rowcolor{nonllmgray}
    WLv3        & 48.63  & 49.80  & 9.44  & 11.07 & 36.68  & 31.12 \\
    Gemma4      & 38.80  & 72.49  & 21.28 & 24.83 & 63.08  & 44.10 \\
    Phi4-MM     & N/A    & 135.82 & 38.58 & 36.42 & 247.56 & 114.60$^{*}$ \\
    AF3         & 104.44 & 133.84 & 78.38 & 83.23 & 152.97 & 110.57 \\
    Q25O        & 74.88  & 184.39 & 67.77 & 64.63 & 132.42 & 104.82 \\
    \rowcolor{q3ored}
    Q3O         & 48.42  & 156.21 & 83.32 & 45.47 & 51.29  & 76.94 \\
    \midrule
    Walrus-CR   & \underline{32.68} & \underline{40.16}  & \underline{21.10} & \underline{24.32} & \underline{58.53}  & \underline{35.36} \\
    Walrus-R    & \wb{16.80} & \wb{22.95} & \wb{4.68} & \wb{4.67} & \wb{25.97} & \wb{15.01} \\
    \bottomrule
  \end{tabular}
  \vspace{-3mm}
\end{table}

\paragraph{Pipelines and training stages.}
Not every pipeline is present in every stage.
Stage~1 uses cleaned real corpora and does not use the synthetic pipelines for code-switching, multi-talker ASR, or closed captioning.
Stage~2 adds the first versions of these pipelines, and it uses only commercially licensed segment libraries for closed captioning.
Stage~3 adds more synthetic multi-talker mixtures, more code-switching and entertainment data, and the remaining closed-captioning libraries.
The dialogue pipeline also changes between stages, because the translation estimator differs.
Sec.~\secref{sec:curriculum} gives the details of each stage.

\begin{table}[!ht]
  \caption{\textbf{MT-ASR Results: LibriMix Clean 2Mix (left) and Japanese Daily Talk (right).} Metrics: cpWER/cpCER and tcpWER/tcpCER ($\downarrow$). tcpWER/tcpCER were calculated with \texttt{equidistant\_interval} option and five-second collar tolerance. Japanese Daily Talk is evaluated on CEJC (Corpus of Everyday Japanese Conversation) on 30-second segments.}
  \label{tab:mt-librimix-jdailytalk}
  \vspace{1mm}
  \centering
  \begin{tabular}{lcccc}
    \toprule
    & \multicolumn{2}{c}{\textbf{LibriMix (Clean 2Mix)}} & \multicolumn{2}{c}{\textbf{Japanese Daily Talk}} \\
    \cmidrule(lr){2-3}\cmidrule(lr){4-5}
    \textbf{Model} & \textbf{cpWER} & \textbf{tcpWER} & \textbf{cpCER} & \textbf{tcpCER} \\
    \midrule
    \rowcolor{nonllmgray}
    VibeVoice-ASR~\cite{peng2026vibevoice}   & 59.34    & 59.36    & 71.99 & 72.17 \\
    Gemma4          & N/A   & N/A   & N/A   & N/A \\
    Phi4-MM         & N/A   & N/A   & N/A   & N/A \\
    AF3              & N/A   & N/A   & N/A   & N/A \\
    Q25O             & N/A   & N/A   & N/A   & N/A \\
    \rowcolor{q3ored}
    Q3O             & 37.04    & N/A   & N/A   & N/A \\
    \midrule
    Walrus-CR       & \underline{26.84} & \underline{27.12} & \underline{66.60} & \underline{77.22}\\
    Walrus-R        & \wb{19.32} & \wb{19.43} & \wb{56.29} & \wb{57.03} \\
    \bottomrule
  \end{tabular}
  \vspace{-3mm}
\end{table}

\begin{table}[!ht]
  \caption{\textbf{AST Results: CoVoST-2, X$\to$En (BLEU $\uparrow$).} *This average score reflects only the supported languages.}
  \label{tab:ast-covost}
  \vspace{1mm}
  \centering
  \begin{tabular}{lccccccccc}
    \toprule
    \textbf{Model} & \textbf{DE} & \textbf{ES} & \textbf{FR} & \textbf{IT} & \textbf{JA} & \textbf{PT} & \textbf{RU} & \textbf{ZH} & \textbf{AVG} \\
    \midrule
    \rowcolor{nonllmgray}
    WLv3        & 36.14 & 37.06 & 33.87 & 35.96 & 23.53 & 53.05 & 41.61 & 11.50 & 34.09 \\
    Gemma4      & 42.01 & 37.15 & 41.04 & \underline{43.02} & 25.50 & 51.12 & 46.44 & 17.90 & 38.02 \\
    Phi4-MM     & 39.81 & \textbf{43.60} & \textbf{42.24} & 41.42 & \underline{30.54} & 55.28 & N/A   & 22.39 & 39.33$^{*}$ \\
    AF3         & 32.80 & 24.19 & 23.34 & 26.68 & 13.08 & 36.46 & 33.49 & 13.24 & 25.41 \\
    Q25O        & 42.42 & 40.51 & 37.46 & 34.95 & 14.57 & \underline{56.65} & 50.44 & \underline{22.58} & 37.44 \\
    \rowcolor{q3ored}
    Q3O         & 46.06 & 43.01 & 41.95 & 42.61 & 30.54 & 59.65 & 54.02 & 28.09 & 43.24 \\
    \midrule
    Walrus-CR   & \underline{43.22} & 41.12 & 40.73 & 41.89 & 28.20 & 54.32 & \underline{52.20} & 20.56 & \underline{40.28} \\
    Walrus-R    & \wb{44.49} & \underline{41.54} & \underline{42.09} & \wb{43.19} & \wb{31.66} & \wb{59.54} & \wb{54.15} & \wb{24.40} & \wb{42.63} \\
    \bottomrule
  \end{tabular}
  \vspace{-3mm}
\end{table}

\section{Breakdown by task and language}
\label{app:breakdowns}
We report here the full per-language and per-dataset breakdowns for the results discussed in Sec.~\secref{sec:results}.
Numbers for prior systems are taken from their official technical reports where available, and otherwise computed with our internal evaluation system.
We include Whisper Large-v3 (\texttt{WLv3}, highlighted in gray) and Qwen3-Omni-30B (\texttt{Q3O}, highlighted in red) as additional reference points. Neither is included in the bold/underline calculations, since \texttt{WLv3} is not a SpeechLLM and \texttt{Q3O} is four times larger than Walrus and any other SpeechLLM included in the comparison.

\subsection{Automatic Speech Recognition}
Tables~\ref{tab:ls}-\ref{tab:asr-internal} report ASR results across all public and internal benchmarks.

\noindent\textbf{LibriSpeech} (Table~\ref{tab:ls}).
\texttt{Walrus-R} reaches 1.65/3.85 WER on clean/other, close to the best comparable system \texttt{AF3} (1.57/3.13).
The gap is small and consistent with the fact that LibriSpeech is a clean, read-speech benchmark that does not exercise the harder skills Walrus is built for.

\noindent\textbf{MLS} (Table~\ref{tab:mls}).
\texttt{Walrus-R} achieves the best average WER of 6.89 across all eight languages, ahead of \texttt{Phi4-MM} (8.20, computed over supported languages only) and \texttt{Gemma4} (9.67).
Gains are largest on French, German, Italian, Polish, and Spanish, reflecting the broad multilingual coverage of our Stage-1 and Stage-2 data.

\noindent\textbf{FLEURS} (Table~\ref{tab:fleurs-asr}).
\texttt{Walrus-R} reaches an average WER/CER of 5.33, second only to \texttt{Phi4-MM} (4.00, computed over supported languages only).
Notable gains appear on Arabic, Hindi, and Korean, where Stage-3 low-resource data has the most impact.

\noindent\textbf{VoxPopuli} (Table~\ref{tab:voxpopuli}).
Walrus leads clearly on this benchmark, with \texttt{Walrus-R} reaching 7.84 WER and \texttt{Walrus-CR} already at 8.38, both well ahead of the best baseline \texttt{Gemma4} (11.78).
The improvement is consistent across all five languages.

\noindent\textbf{CoVoST-2 ASR} (Table~\ref{tab:covost-asr}).
\texttt{Walrus-R} reaches an average of 6.38 WER/CER, the best among comparable systems.
Gains are particularly strong on Japanese and Chinese, where our multilingual curriculum provides substantial coverage.

\noindent\textbf{Internal entertainment ASR} (Table~\ref{tab:asr-internal}).
This is the benchmark where Stage-3 specialization has the clearest impact.
\texttt{Walrus-R} achieves the best result in every language, with the largest gains in Japanese (11.97 vs.\ 23.69 for \texttt{Walrus-CR}), French (24.55 vs.\ 59.16), and German (27.68 vs.\ 39.71).

\begin{table}[!ht]
  \caption{\textbf{AST Results: FLEURS, X$\to$En (top) and En$\to$X (bottom), BLEU ($\uparrow$).} *This average score reflects only the supported languages.}
  \label{tab:ast-fleurs-xen-enx}
  \vspace{1mm}
  \centering
  \resizebox{\columnwidth}{!}{%
  \begin{tabular}{lccccccccccc}
    \toprule
    \textbf{Model} & \textbf{AR} & \textbf{DE} & \textbf{ES} & \textbf{FR} & \textbf{HI} & \textbf{IT} & \textbf{JA} & \textbf{KO} & \textbf{PT} & \textbf{ZH} & \textbf{AVG} \\
    \midrule
    \multicolumn{12}{@{}c}{\textit{X$\to$En}} \\
    \midrule
    \rowcolor{nonllmgray}
    WLv3        & 20.63 & 33.27 & 22.27 & 31.67 & 19.40 & 23.57 & 16.51 & 19.05 & 36.68 & 15.80 & 23.89 \\
    Gemma4      & 29.56 & 33.55 & 23.11 & 31.22 & 27.46 & 23.74 & 22.47 & \underline{23.44} & 36.79 & 21.63 & 27.29 \\
    Phi4-MM     & N/A   & \underline{37.71} & 25.33 & 35.10 & N/A   & 26.06 & 21.62 & N/A   & 40.80 & \underline{22.37} & 29.86$^{*}$ \\
    AF3         & 14.05 & 33.27 & 23.43 & 28.83 & 8.62  & 23.08 & 11.90 & 14.32 & 35.39 & 16.43 & 20.93 \\
    Q25O        & 18.79 & 35.30 & 24.81 & 31.68 & 16.09 & 23.94 & 18.36 & 19.01 & \underline{41.48} & 19.43 & 24.89 \\
    \rowcolor{q3ored}
    Q3O         & 33.31 & 44.42 & 30.40 & 42.10 & 32.24 & 31.30 & 26.41 & 29.80 & 47.32 & 29.20 & 34.61 \\
    \midrule
    Walrus-CR   & \underline{35.89} & \underline{37.71} & \underline{26.32} & \underline{36.31} & \underline{31.77} & \underline{27.40} & \underline{25.36} & 20.07 & 40.95 & 20.33 & \underline{30.21} \\
    Walrus-R    & \wb{38.16} & \wb{38.91} & \wb{27.53} & \wb{38.13} & \wb{37.36} & \wb{29.28} & \wb{27.01} & \wb{24.23} & \wb{42.71} & \wb{27.22} & \wb{33.05} \\
    \midrule \midrule
    \multicolumn{12}{@{}c}{\textit{En$\to$X}} \\
    \midrule
    \rowcolor{nonllmgray}
    WLv3        & N/A   & N/A   & N/A   & N/A   & N/A   & N/A   & N/A   & N/A   & N/A   & N/A   & N/A \\
    Gemma4      & 16.55 & 24.49 & 22.66 & 37.20 & 22.26 & 23.90 & 39.08 & \textbf{36.91} & 37.30 & \underline{34.25} & 29.46 \\
    Phi4-MM     & N/A   & \underline{34.44} & 23.66 & 37.62 & N/A   & 23.44 & 30.67 & N/A   & 37.79 & \textbf{37.10} & 32.15$^{*}$ \\
    AF3         & N/A   & N/A   & N/A   & N/A   & N/A   & N/A   & N/A   & N/A   & N/A   & N/A   & N/A \\
    Q25O        & 12.65 & 25.74 & 20.73 & 30.99 & 18.84 & 14.54 & 10.05 & 11.78 & 29.77 & 15.15 & 19.02 \\
    \rowcolor{q3ored}
    Q3O         & 28.18 & 35.77 & 26.00 & 48.97 & 46.86 & 29.03 & 14.50 & 13.09 & 45.09 & 17.79 & 29.75 \\
    \midrule
    Walrus-CR   & \underline{26.33} & 34.06 & \underline{25.04} & \underline{44.22} & \underline{25.32} & \underline{28.09} & \underline{40.02} & 32.18 & \underline{38.58} & 30.41 & \underline{32.42} \\
    Walrus-R    & \wb{27.69} & \wb{34.85} & \wb{25.87} & \wb{49.77} & \wb{29.44} & \wb{29.82} & \wb{40.04} & \underline{34.23} & \wb{39.57} & 32.01 & \wb{34.33} \\
    \bottomrule
  \end{tabular}
  }
  \vspace{-3mm}
\end{table}

\subsection{Code-Switching ASR}
Table~\ref{tab:cs-read} reports CS-ASR results on the CS-FLEURS Read and XTTS test sets, which are OOD for Walrus.
\texttt{Walrus-R} is the strongest system by a large margin on both sets, reaching 7.23 CER on Read and 15.01 on XTTS, against 32.30/44.10 for the best baseline (\texttt{Gemma4}) and 26.69/35.36 for \texttt{Walrus-CR}.
The gain from \texttt{Walrus-CR} to \texttt{Walrus-R} is the largest single improvement across all benchmarks, and it comes entirely from the synthetic CS-ASR pipeline introduced in Stage~3.
Performance is consistent across all five language pairs, with the largest absolute gains on English-Japanese and English-Hindi.

\subsection{Multi-Talker ASR}
Table~\ref{tab:mt-librimix-jdailytalk} reports MT-ASR results on LibriMix Clean 2-Mix and the Japanese Daily Talk corpus (CEJC).
All comparable baselines produce no valid output on these benchmarks, as they are not designed for overlapping-speech scenarios.
\texttt{Walrus-R} reaches 19.32 cpWER on LibriMix and 56.29 cpCER on Japanese Daily Talk, with \texttt{Walrus-CR} already competitive at 26.84 and 66.60, respectively. 

\subsection{Speech Translation}

\noindent\textbf{AST CoVoST-2 X$\to$En} (Table~\ref{tab:ast-covost}).
\texttt{Walrus-R} achieves the best average BLEU of 42.63, ahead of \texttt{Phi4-MM} (39.33) and \texttt{Gemma4} (38.02).
Gains are consistent across languages, with the strongest improvements on Japanese, Russian, and Chinese.

\noindent\textbf{AST FLEURS X$\to$En and En$\to$X} (Table~\ref{tab:ast-fleurs-xen-enx}).
On the OOD FLEURS X$\to$En, \texttt{Walrus-R} reaches 33.05 BLEU, again ahead of \texttt{Phi4-MM} (29.86) and \texttt{Gemma4} (27.29).
On En$\to$X, \texttt{Walrus-R} achieves 34.33 BLEU, the best among all comparable systems, with strong performance on Fr, De, and Ja.

\subsection{Speech-to-Speech Translation}

\noindent\textbf{S2ST CVSS-C X$\to$En} (Table~\ref{tab:s2st-cvss}).
\texttt{Walrus-R} reaches 43.63 ASR-BLEU, the best result among all systems, including the $4\times$ larger Qwen3-Omni-30B (41.27).
\texttt{Walrus-CR} is also competitive at 43.27, showing that S2ST capability is already strong after Stage~2.

\noindent\textbf{S2ST FLEURS X$\to$En and En$\to$X} (Table~\ref{tab:s2st-fleurs-xen}).
On X$\to$En, \texttt{Walrus-R} achieves 32.28 ASR-BLEU, ahead of \texttt{Q25O} (25.15) and close to the bigger \texttt{Q3O} (33.60).
On En$\to$X, \texttt{Walrus} is the only system among comparable models that supports this direction, reaching 22.94 ASR-BLEU (research variant).

\begin{table}[!ht]
  \caption{\textbf{S2ST Results: CVSS-C, X$\to$En (ASR-BLEU $\uparrow$).}}
  \label{tab:s2st-cvss}
  \vspace{1mm}
  \centering
  \begin{tabular}{lccccccccc}
    \toprule
    \textbf{Model} & \textbf{DE} & \textbf{ES} & \textbf{FR} & \textbf{IT} & \textbf{JA} & \textbf{PT} & \textbf{RU} & \textbf{ZH} & \textbf{AVG} \\
    \midrule
    Gemma4      & N/A   & N/A   & N/A   & N/A   & N/A   & N/A   & N/A   & N/A   & N/A \\
    Phi4-MM     & N/A   & N/A   & N/A   & N/A   & N/A   & N/A   & N/A   & N/A   & N/A \\
    AF3         & N/A   & N/A   & N/A   & N/A   & N/A   & N/A   & N/A   & N/A   & N/A \\
    Q25O        & 40.39 & 38.32 & 35.54 & 33.67 & 15.61 & 55.10 & 48.15 & 21.39 & 36.02 \\
    \rowcolor{q3ored}
    Q3O         & 44.10 & 41.32 & 38.30 & 40.21 & 30.04 & 58.12 & 52.19 & 25.90 & 41.27 \\
    \midrule
    Walrus-CR   & \wb{49.21} & \underline{47.12} & \wb{40.73} & \wb{41.89} & \underline{32.14} & \underline{57.33} & \wb{52.58} & \underline{25.16} & \underline{43.27} \\
    Walrus-R    & \underline{49.01} & \wb{48.97} & \underline{37.96} & \underline{40.27} & \wb{33.85} & \wb{59.99} & \underline{52.51} & \wb{26.44} & \wb{43.63} \\
    \bottomrule
  \end{tabular}
  \vspace{-3mm}
\end{table}

\begin{table}[!ht]
  \caption{\textbf{S2ST Results: FLEURS, X$\to$En (top) and En$\to$X (bottom), ASR-BLEU ($\uparrow$).}}
  \label{tab:s2st-fleurs-xen}
  \vspace{1mm}
  \centering
  \resizebox{\columnwidth}{!}{%
  \begin{tabular}{lccccccccccc}
    \toprule
    \textbf{Model} & \textbf{AR} & \textbf{DE} & \textbf{ES} & \textbf{FR} & \textbf{HI} & \textbf{IT} & \textbf{JA} & \textbf{KO} & \textbf{PT} & \textbf{ZH} & \textbf{AVG} \\
    \midrule
    \multicolumn{12}{@{}c}{\textit{X$\to$En}} \\
    \midrule
    Gemma4      & N/A   & N/A   & N/A   & N/A   & N/A   & N/A   & N/A   & N/A   & N/A   & N/A   & N/A \\
    Phi4-MM     & N/A   & N/A   & N/A   & N/A   & N/A   & N/A   & N/A   & N/A   & N/A   & N/A   & N/A \\
    AF3         & N/A   & N/A   & N/A   & N/A   & N/A   & N/A   & N/A   & N/A   & N/A   & N/A   & N/A \\
    Q25O        & 20.63 & \underline{36.40} & 23.71 & 31.54 & 15.78 & 25.04 & 17.77 & 20.39 & \underline{38.91} & 21.33 & 25.15 \\
    \rowcolor{q3ored}
    Q3O         & 32.80 & 43.10 & 29.30 & 41.00 & 31.52 & 30.34 & 25.61 & 29.30 & 45.41 & 28.30 & 33.60 \\
    \midrule
    Walrus-CR   & \underline{34.23} & 35.00 & \underline{25.86} & \underline{33.79} & \underline{27.94} & \underline{25.47} & \underline{22.92} & \underline{22.79} & 36.17 & \underline{21.71} & \underline{28.59} \\
    Walrus-R    & \wb{38.23} & \wb{38.87} & \wb{27.90} & \wb{38.12} & \wb{32.75} & \wb{28.62} & \wb{26.51} & \wb{25.38} & \wb{40.43} & \wb{25.98} & \wb{32.28} \\
    \midrule \midrule
    \multicolumn{12}{@{}c}{\textit{En$\to$X}} \\
    \midrule
    Gemma4      & N/A   & N/A   & N/A   & N/A   & N/A   & N/A   & N/A   & N/A   & N/A   & N/A   & N/A \\
    Phi4-MM     & N/A   & N/A   & N/A   & N/A   & N/A   & N/A   & N/A   & N/A   & N/A   & N/A   & N/A \\
    AF3         & N/A   & N/A   & N/A   & N/A   & N/A   & N/A   & N/A   & N/A   & N/A   & N/A   & N/A \\
    Q25O        & N/A   & N/A   & N/A   & N/A   & N/A   & N/A   & N/A   & N/A   & N/A   & N/A   & N/A \\
    \rowcolor{q3ored}
    Q3O         & 17.83 & 32.10 & 25.11 & 43.89 & 23.59 & 25.40 & 13.04 & 11.13 & 40.29 & 24.29 & 25.67 \\
    \midrule
    Walrus-CR   & \underline{12.41} & \underline{24.14} & \underline{18.20} & \underline{30.44} & \underline{11.03} & \underline{12.98} & \underline{10.41} & \underline{7.55}  & \underline{32.75} & \underline{12.04} & \underline{17.19} \\
    Walrus-R    & \wb{15.01} & \wb{39.03} & \wb{22.55} & \wb{36.07} & \wb{18.40} & \wb{20.80} & \wb{11.75} & \wb{11.14} & \wb{38.50} & \wb{16.10} & \wb{22.94} \\
    \bottomrule
  \end{tabular}
  }
  \vspace{-3mm}
\end{table}

\begin{table}[ht]
    \caption{\textbf{Speech-to-speech translation on FLEURS, X$\to$En (left) and En$\to$X (right).}
    Metrics are ASR-BLEU, UTMOS (predicted naturalness), speaker similarity (SSim), emotion similarity (ESim), and AutoPCP (APCP, predicted prosody similarity).
    Higher is better for all metrics, and scores are averages over languages.
    N/A means that the system does not support the direction.
    Qwen3-Omni (Q3O, shaded) is much larger and is not ranked.
    Among the other systems, the best score is in bold and the second best is underlined.}
    \label{tab:utmos-ssim-apcp}
    \vspace{1mm}
    \centering
    \resizebox{\columnwidth}{!}{%
    \begin{tabular}{lcccccccccc} 
    \toprule
    & \multicolumn{5}{c}{\textbf{X$\to$En}} & \multicolumn{5}{c}{\textbf{En$\to$X}} \\
    \cmidrule(lr){2-6}\cmidrule(lr){7-11}
    \textbf{Model} & \textbf{ASR-BLEU} & \textbf{UTMOS} & \textbf{SSim} & \textbf{ESim} & \textbf{APCP} & \textbf{ASR-BLEU} & \textbf{UTMOS} & \textbf{SSim} & \textbf{ESim} & \textbf{APCP} \\ 
    \midrule
    Kraken~\cite{futami2026kraken}      & \textbf{32.30}    & 3.13             & \underline{0.75} & \underline{0.77} & 2.48             
                & N/A  & N/A  & N/A  & N/A  & N/A \\
    Seamless~\cite{barrault2023seamless}    & 32.10             & 3.06             & 0.66             & 0.76             & 2.05             
                & \textbf{27.33}  & 2.82  & 0.58  & 0.56  & 2.25 \\
    Q2.5O       & 25.10             & \textbf{3.48}    & 0.49             & 0.53             & 2.40             
                & N/A  & N/A  & N/A  & N/A  & N/A \\
    \rowcolor{q3ored}
    Q3O         & 33.60             & 3.59             & 0.61             & 0.75             & 2.19             
                & 25.67  & 3.27  & 0.49  & 0.44  & 2.12 \\ 
    \midrule
    Walrus-CR   & 28.59             & 3.17             & \underline{0.75} & 0.76             & \underline{2.54} 
                & 17.19  & \underline{3.01}  & \underline{0.67}  & \underline{0.69}  & \underline{2.41}  \\ 
    Walrus-R    & \underline{32.28} & \underline{3.19} & \wb{0.77}        & \wb{0.81}        & \wb{2.57}    
                & \underline{22.94}  & \wb{3.05}  & \wb{0.70}  & \wb{0.73}  & \wb{2.45}  \\ 
    \bottomrule
    \end{tabular}
    }
    \vspace{-3mm}
\end{table}

\noindent\textbf{S2ST quality beyond ASR-BLEU} (Table~\ref{tab:utmos-ssim-apcp}).
ASR-BLEU measures translation accuracy but does not capture the full quality of spoken output.
A complete evaluation of S2ST must also consider speaker consistency, emotion preservation, prosody preservation, and naturalness, as detailed in~\cite{koudounas2026benchmarking}.
We thus compute several metrics, using the COMPASS toolkit~\cite{koudounas2026benchmarking}:
UTMOS-v2 (\texttt{UTMOS})\footnote{\href{https://github.com/sarulab-speech/UTMOSv2}{\texttt{github.com/sarulab-speech/UTMOSv2}}} predicts the naturalness of the output speech; speaker similarity (\texttt{SSim}) is the similarity between the speaker of the source speech and the speaker of the output, measured via WavLM-SV;\footnote{\href{https://github.com/sarulab-speech/UTMOSv2}{\texttt{huggingface.co/microsoft/wavlm-base-sv}}} emotion similarity (\texttt{ESim}) is the similarity between the emotion of the source and of the output measured using Emotion2Vec;\footnote{\href{https://huggingface.co/emotion2vec/emotion2vec_plus_base}{\texttt{huggingface.co/emotion2vec/emotion2vec\_plus\_base}}} AutoPCP (\texttt{APCP}) predicts how similar the prosody of the output is to the prosody of the source~\cite{barrault2023seamless}.
We compare Walrus with four systems: Qwen2.5/3-Omni, Kraken~\cite{futami2026kraken}, the system whose speech decoder Walrus adopts, and Seamless~\cite{barrault2023seamless}, a dedicated multilingual S2ST system.
Q2.5O and Kraken support only X$\to$En translation.

On X$\to$En, \texttt{Walrus-R} reaches 32.28 ASR-BLEU. This is on par with Kraken (32.30) and Seamless (32.10), 7.18 points above \texttt{Q2.5O} (25.10), but 1.32 points lower than \texttt{Q3O} (33.60).
On En$\to$X, \texttt{Walrus-R} achieves 22.94 ASR-BLEU, slightly below both Seamless and \texttt{Q3O}.

The strongest result of Walrus is the preservation of the speaker, emotion, and prosody.
On X$\to$En, \texttt{Walrus-R} has the best \texttt{SSim} (0.77), \texttt{ESim} (0.81), and AutoPCP (2.57) of all systems, including \texttt{Q3O} (0.61, 0.75, and 2.19).
Compared with Seamless, it improves \texttt{SSim} by 0.11, \texttt{ESim} by 0.05, and AutoPCP by 0.52.
Compared with Kraken, which shares our speech decoder, the gains are smaller (0.02, 0.04, and 0.09).
On En$\to$X, the gaps are larger: \texttt{Walrus-R} reaches 0.70 \texttt{SSim}, 0.73 \texttt{ESim}, and 2.45 AutoPCP, compared to 0.58/0.56/2.25 for Seamless and 0.49/0.44/2.12 for \texttt{Q3O}.
Walrus conditions its vocoder on the source speech (Sec.~\secref{sec:model}), which is consistent with this result.

The naturalness scores are more mixed.
\texttt{Walrus-R} has a higher \texttt{UTMOS} than Seamless in both directions (3.19 vs 3.06 and 3.05 vs 2.82).
The Qwen Omni models score higher on X$\to$En (3.48 for Q2.5O and 3.59 for Q3O), and \texttt{Q3O} also scores higher on En$\to$X (3.27).
Both these models generate speech with fixed reference voices: this can make the speech cleaner and more natural, but it does not preserve the prosody and emotion of the source, which fits their low \texttt{ESim} and AutoPCP scores.
Walrus, in contrast, conditions its vocoder on the source speech; FLEURS contains recordings of uneven quality, so this conditioning can carry the noise of the source into the output and eventually lower the \texttt{UTMOS}.

\texttt{Walrus-R} improves on \texttt{Walrus-CR} on all columns of Table~\ref{tab:utmos-ssim-apcp}.
The largest gains are in ASR-BLEU, with +3.69 on X$\to$En (28.59 to 32.28) and +5.75 on En$\to$X (17.19 to 22.94), while the expressivity metrics improve by smaller amounts: \texttt{SSim} improves by 0.02 and 0.03, \texttt{ESim} by 0.05 and 0.04, and AutoPCP by 0.03 and 0.04.
Stage 3 adds S2ST data and raises the S2ST task weight from 0.5 to 2.0 (Sec.~\secref{sec:stage3}): translation quality benefits most, while the voice preservation of \texttt{Walrus-CR} is already close to that of \texttt{Walrus-R}.

\subsection{Audio Understanding Benchmarks}
\texttt{Walrus-R} achieves 71.92\% accuracy on MMAU-Pro, far ahead of the best comparable baseline \texttt{Q25O} (52.20\%) and above commercial systems such as \texttt{Gemini-Flash-3.7} (70.40\%) and \texttt{GPT4o-Audio} (52.50\%).\footnote{See official leaderboard here: \href{https://sonalkum.github.io/mmau-pro/}{sonalkum.github.io/mmau-pro}.}
On MMSU, \texttt{Walrus-R} reaches 64.14\%, ahead of \texttt{AF3} (61.40\%) and \texttt{Q25O} (60.57\%).
The gap between \texttt{Walrus-CR} (50.18\% / 55.18\%) and \texttt{Walrus-R} on both benchmarks is large and explained by the AudioReason-10M data introduced in Stage~3.


\begin{table}[!ht]
    \centering
    \caption{\textbf{Primary (most frequently sampled) prompt for each task.} During training, each prompt is drawn uniformly from a pool of 10-20 paraphrases with the listed prompt oversampled; slot variables (\texttt{\{lang1\}}, \texttt{\{lang2\}}, \texttt{\{context\}}, \texttt{\{keywords\}}, \texttt{\{labels\}}, \texttt{\{speaker\}}, \texttt{\{start\}}, \texttt{\{end\}}, \texttt{\{agent\}}, \texttt{\{user\}}) are filled at data-construction time.}
    \label{tab:prompts}
    \vspace{1mm}
    \small
    \renewcommand{\arraystretch}{1.15}
    \begin{tabular}{@{}p{0.34\linewidth} p{0.60\linewidth}@{}}
    \toprule
    \textbf{Task} & \textbf{Primary prompt} \\
    \midrule
    \rowcolor{stdblue}
    \multicolumn{2}{@{}c}{\textit{Recognition}} \\
    \midrule
    ASR & Transcribe the \{lang1\} speech: \\
    Context-ASR & Transcribe the \{lang1\} speech from the context \{context\}: \\
    Keyword-biased ASR & Transcribe the \{lang1\} speech, biasing toward these keywords: \{keywords\}: \\
    Code-switched ASR & Transcribe the code-switched speech in \{lang1\} and \{lang2\}: \\
    Dialogue-contextual ASR & Given the previous conversation between \{agent\} and \{user\}: \{context\} Transcribe the \{lang1\} speech: \\
    Timestamped ASR & Transcribe the \{language\} utterance with timestamps: \\
    Multi-talker ASR & Provide a speaker-attributed transcription of this multi-talker \{language\} audio: \\
    Timestamped multi-talker ASR & Provide a timestamped, speaker-attributed transcription of this \{language\} audio: \\
    Multi-talker ASR (CoT) & First, count the speakers in this mixture. Then, transcribe each speaker: \\
    Timestamped multi-talker ASR (CoT) & First, estimate how many speakers are in this \{language\} mixture. Then, provide a timestamped transcription of each speaker: \\
    Timestamp QA & What is the speaker saying between seconds \{start\} and \{end\}? \\
    Target-speaker ASR & Please provide the transcription for \{speaker\} only: \\
    \midrule
    \rowcolor{specgreen}
    \multicolumn{2}{@{}c}{\textit{Translation}} \\
    \midrule
    MT & Translate \{lang1\} text into \{lang2\} text: \\
    AST (direct) & Translate the \{lang1\} speech into \{lang2\} text: \\
    AST (cascaded) & Transcribe the \{lang1\} speech and then translate into \{lang2\} text: \\
    Context-AST (direct) & Translate the \{lang1\} speech into \{lang2\} text, with context '\{context\}': \\
    Context-AST (cascaded) & Transcribe the \{lang1\} speech and then translate it into \{lang2\} text, with context \{context\}: \\
    S2ST (direct) & Translate the \{lang1\} speech into \{lang2\} speech: \\
    S2ST (cascaded) & Transcribe \{lang1\} speech and then translate into \{lang2\} text and then into speech: \\
    Context-S2ST (direct) & Translate the \{lang1\} speech into \{lang2\} speech, with context \{context\}: \\
    Context-S2ST (cascaded) & Transcribe the \{lang1\} speech, translate it into \{lang2\} text, and then into speech, with context '\{context\}': \\
    \midrule
    \rowcolor{envred}
    \multicolumn{2}{@{}c}{\textit{Generation}} \\
    \midrule
    TTS & Generate \{lang1\} speech of the text: \\
    \midrule
    \rowcolor{accentyellow}
    \multicolumn{2}{@{}c}{\textit{Understanding}} \\
    \midrule
    SER & Recognize the emotion in the speech; choose one from [\{labels\}]: \\
    Audio captioning & Generate a caption for this speech: \\
    CC [audio-event ASR] & Transcribe the \{lang1\} speech and include any background audio events or sounds: \\
    CC [event sequence] & Identify the structural timeline of this audio (e.g., [Speech], [Sigh], [Cat meowing]): \\
    CC [timestamped event ASR] & Provide a timestamped transcription of the \{lang1\} speech along with time-aligned sound events: \\
    CC [diarized-event ASR] & Provide a timestamped diarized transcription in \{lang1\}, including background audio events: \\
    CC [full scene description] & Summarize the entire audio scene (total events, speakers, SFX, music, etc.) and provide a detailed timeline breakdown: \\
    \bottomrule
    \end{tabular}
\end{table}

\end{document}